\documentclass[12pt, draftclsnofoot, onecolumn]{IEEEtran}
\usepackage{amsthm}
\usepackage{array}
\DeclareMathAlphabet{\mathpzc}{OT1}{pzc}{m}{it}
\usepackage{mathrsfs}
\usepackage{amssymb}
\usepackage{color}
\usepackage{amsmath}
\usepackage{amsfonts}
\usepackage{graphicx}
\usepackage{algorithm}
\usepackage{algorithmic}
\usepackage{epstopdf}
\usepackage{cite}
\usepackage{hyperref}
\usepackage{subfigure}
\usepackage{bm}
\usepackage{fancyhdr}
\usepackage{etoolbox}
\usepackage{multirow}
\usepackage{setspace}
\usepackage{booktabs}
\makeatletter
\patchcmd{\@makecaption}
{\scshape}
{}
{}
{}
\makeatletter
\patchcmd{\@makecaption} {\\}
{.\ }
{}
{}
\makeatother

\newtheorem{theorem}{\textbf{Theorem}}

\newtheorem{corollary}{\textbf{Corollary}}

\newtheorem{definition}{\textbf{Definition}}

\newtheorem{lemma}{\textbf{Lemma}}

\newtheorem{remark}{\textbf{Remark}}

\begin{document}
\begin{spacing}{1.0}
\abovedisplayskip=2pt
\belowdisplayskip=2pt
\title{Exploiting Computation Replication for Mobile Edge Computing: A Fundamental Computation-Communication Tradeoff Study}
\author{Kuikui Li, Meixia Tao,~\IEEEmembership{Fellow,~IEEE,} Zhiyong Chen,~\IEEEmembership{Member,~IEEE}
\thanks{This work is supported by the NSF of China under grant 61941106, 61671291, and the National Key R$\&$D Project of China under grant 2019YFB1802701 and 2019YFB1802702. Part of this work was presented in IEEE GLOBECOM 2018 \cite{gckkl}, and part was presented in IEEE ISIT 2019 \cite{kklisit}.

The authors are with Shanghai Institute for Advanced Communication and Data Science, and Department of Electrical Engineering, Shanghai Jiao Tong University, Shanghai 200240, China. Z. Chen is also with the Cooperative Medianet Innovation
Center, Shanghai Jiao Tong University, Shanghai 200240, China. E-mail: \{kuikuili, mxtao, zhiyongchen\}@sjtu.edu.cn. The corresponding author is Meixia Tao.}}
\maketitle
\begin{abstract}
Existing works on task offloading in mobile edge computing (MEC) networks often assume a task is executed once at a single edge node (EN). Downloading the computed result from the EN back to the mobile user may suffer long delay if the downlink channel experiences strong interference or deep fading. This paper exploits the idea of computation replication in MEC networks to speed up the downloading phase. Computation replication allows each user to offload its task to multiple ENs for repetitive execution so as to create multiple copies of the computed result at different ENs which can then enable transmission cooperation and hence reduce the communication latency for result downloading. Yet, computation replication may also increase the communication latency for task uploading, despite the obvious increase in computation load.  The main contribution of this work is to characterize asymptotically an order-optimal upload-download communication latency pair for a given computation load in a multi-user multi-server MEC network. Analysis shows when the computation load increases within a certain range, the downloading time decreases in an inversely proportional way if it is binary offloading or decreases linearly if it is partial offloading, both at the expense of linear increase in the uploading time.
\end{abstract}
\section{Introduction}
The explosive growth of Internet of Things is driving the emergence of new mobile applications that demand intensive computation and stringent latency, such as intelligent navigation, online gaming, virtual reality (VR), and augmented reality (AR)\cite{7498684}. The limited computation and energy resources at mobile devices pose a great challenge for supporting these new applications\cite{Miettinen2010Energy}. Mobile edge computing (MEC) is envisioned as a promising network architecture to address this challenge by providing cloud-computing services at the edge nodes (ENs) of mobile networks, such as wireless access points (APs) and base stations (BSs)\cite{MEC,The_Case}. By offloading computation-intensive tasks from mobile users to their nearby server-enabled ENs for processing, MEC systems have great potential to prolong the battery lifetime of mobile devices and reduce the overall task execution latency.

Task offloading and resource allocation are crucial problems in MEC systems. This is because offloading a task from a user to its associated EN involves extra overhead in transmission energy and communication latency due to the input data uploading and computation result downloading\cite{6923537,6189008,7541539}. In the existing literature, the task offloading and resource (both radio resource and computation resource) allocation problems have been studied for single-user single-server MEC systems in \cite{StochasticChannel,tvtMEC}, for multi-user single-server MEC systems in \cite{K.Huang,8234686,7956189,Simeone,8334188,8387798,Mobile-EdgeCloud,SSardellitti}, for single-user multi-server MEC systems in \cite{Quek}, and for multi-user multi-server MEC systems in\cite{muluserserver}. These problems are often formulated as minimizing the energy consumption under latency constraints or minimizing the latency subject to energy constraints, so as to strike a good balance between communication efficiency and computation efficiency. The task offloading strategy also depends on whether the task is dividable, known as partial offloading \cite{tvtMEC,K.Huang,8234686,7956189,Simeone}, or has to be executed as a whole, known as binary offloading \cite{8334188,8387798,Mobile-EdgeCloud,SSardellitti,Quek,muluserserver}.

Note that in the aforementioned literature \cite{StochasticChannel,tvtMEC,K.Huang,8234686,7956189,Simeone,8334188,8387798,Mobile-EdgeCloud,SSardellitti,Quek,muluserserver}, each task or subtask is executed once at one EN only. As a result, the computed result of the task/subtask being offloaded is only available at one EN without any diversity. Consider the scenario where the downlink channel from the EN to the user experiences deep fading or suffers from strong interference, downloading the computed results back to the user may incur very low transmission rate and consequently cause significantly long delay in completing the overall task. This work aims to address this issue by  exploiting computation replication when there are multiple edge servers.

The main idea of computation replication is to let mobile users offload their tasks or subtasks to multiple ENs for repeated execution so as to create multiple copies of the computed result at different ENs which then can enable the multiple ENs to cooperatively transmit the computed result back to users in the downlink. This transmission cooperation can mitigate interferences across users, overcome deep fading, and hence increase the transmission rate of downlink channels. Thus, the downloading time can be reduced. But note that a side effect of replicating tasks on multiple ENs is that it introduces more data traffic in the uplink and hence may increase the uploading time compared with offloading to a single EN. That means computation replication can induce a tradeoff between uploading time and downloading time. Take a 3-user 3-server MEC network for example. First, consider that each user offloads its task to all $3$ servers for repeated computing. In the uploading phase, each user takes turn to multicast its task input to all $3$ ENs, consuming 3 time slots in total. In the downloading phase, all the ENs have the same message to transmit, and the downlink channel thus becomes a virtual MISO broadcast channel, and all computed results can be delivered to users within $1$ time slot by using zero-forcing precoding\cite{coBC}. For the baseline scheme where each task is offloaded to an individual EN, both the uplink and downlink channels are $3$-user interference channels\cite{K_dof}. Both task uploading and downloading can be completed within $2$ time slots by using interference alignment. It is seen that by computation replication, the downloading time is reduced from $2$ time slots to $1$ time slot while the uploading time is increased from $2$ time slots to $3$ time slots. For those computation tasks whose output data size is much larger than the input data size or the input data size is negligible, computation replication can bring significant benefits in reducing the overall communication latency. In general, computation replication should be carefully chosen by considering the balance between upload and download times, despite the obvious increase in the computation load.

Computation replication, i.e., replicating a task on multiple servers, has already demonstrated significant advantages in modern computer systems. It can mitigate the random server straggling and thus reduce the task service delay in queuing systems \cite{Speeding_Up,replication,Replications}. It can also create coded multicasting for data shuffling and thus reduce the communication load in distributed computing frameworks, like MapReduce and Spark \cite{fundamental_tradeoff,Scalable_Framework}. Our work is an attempt to exploit this idea of computation replication in MEC systems to enable the transmission cooperation for speeding up the computed result downloading. To our best knowledge, the only work that uses the similar idea of computation replication to enable transmission cooperation for computed result downloading in MEC systems is \cite{8007057}. However, \cite{8007057} relies on a strong assumption that the computation functions are linear so that the tasks can be executed on some linear combinations of the inputs, i.e., coded input, and then the computed coded outputs on all ENs are utilized to zero-force the downlink interference at each user. In addition, the work \cite{8007057} ignores the task uploading phase.

In this work, we exploit computation replication in a general multi-user multi-server MEC system with general computation function. Unlike \cite{8007057}, the transmission cooperation in the downlink enabled by computation replication does not assume linearity of the computation function. It relies purely on the replication of the computed result of each individual task at multiple ENs. Moreover, we consider both task uploading and results downloading and adopt the upload-download latency pair as the performance metric. In specific, we consider an MEC network, where a set of $N$ mobile users offload their tasks to a set of $M$ computing-enabled ENs. Each task has an input data and an output data. We define the \emph{computation load} $r$ as the average number of ENs to compute a task, i.e., the degrees of replication for executing a task. The \emph{communication latency} is defined as the upload-download time pair, denoted as $(\tau^u, \tau^d)$. A fundamental question we would like to address is:
\emph{Given a computation load $r$, what is the minimum achievable communication latency boundary $\left(\tau^u(r),\tau^d(r)\right)$?}

Our work attempts to address the above question for a general MEC network with any $M$ ($\ge2$) ENs and any $N$ ($\ge2$) users, in both binary and partial offloading cases from an information-theoretic perspective. We reveal a fundamental tradeoff between computation load and communication latency, and also present the uploading time and downloading time tradeoff.

We first consider binary offloading where the computation tasks are not dividable.
We propose a task assignment scheme where each task is offloaded to $r$ different ENs for repeated computing, and each EN has an even assignment of $\frac{Nr}{M}$ tasks. By utilizing the duplicated computation results on multiple ENs, transmission cooperation in the form of interference neutralization can be exploited in the data downloading phase. We characterize the communication latency by the pair of the normalized uploading time (NULT) $\tau^u(r)$ and normalized downloading time (NDLT) $\tau^d(r)$. The main distinction in the communication latency $\left(\tau^u(r),\tau^d(r)\right)$ analysis lies at the degree of freedom (DoF) analysis of the so-called circular cooperative interference-multicast channels. We obtain the optimal per-receiver DoF for the uplink channel, and an order-optimal per-receiver DoF for the downlink channel.
Based on these DoF regions, we then develop an order-optimal achievable communication latency pair at any integer computation load. In particular, the NULT is exactly optimal and the NDLT is within a multiplicative gap of 2 to the optimum. We show that the NDLT is an \emph{\textbf{inversely proportional function}} of the computation load in the interval $\frac{M\!N}{M\!+\!N}\!\!\le\!r\!\le\!\!M$, which presents the \emph{\textbf{computation-communication tradeoff}}. We also reveal that the decrease of NDLT is at the expense of increasing the NULT linearly, which forms another NULT-NDLT tradeoff. Part of this result is presented in IEEE ISIT 2019\cite{kklisit}.

Next, we consider partial offloading where each task can be divided arbitrarily. Motivated by the equal file splitting and placement strategy in caching networks (e.g. \cite{7857805,FanXu}), we propose a task partition scheme where the task generated by each user is partitioned into $\binom{M}{r}$ subtasks, and each is offloaded to a distinct subset of $r$ ENs chosen from the total $M$ ENs for repeated computing. By doing so, the uplink is formed as the X-multicast channel whose optimal per-receiver DoF is obtained. The downlink is the cooperative X channel, and an achievable per-receiver DoF is derived with order optimality. We thus develop an order-optimal achievable communication latency pair at any given computation load, and both the achievable NULT and NDLT are within multiplicative gaps of $2$ to their lower bounds. Moreover, the NDLT \emph{\textbf{decreases linearly}} with the computation load in the interval $1\!\le\!r\!\le\!\min\{\!M,\!N\}$, which is also at the expense of increasing the NULT linearly. The similar partial task offloading scheme is presented in our prior work \cite{gckkl} whose focus is to minimize the total communication and computation time for partial offloading but not the fundamental tradeoff between computation load and communication latency. This paper extends the results in \cite{gckkl} by using the new performance metrics to show the computation-communication tradeoff, and also proves the converse for partial offloading, including the lower bounds and multiplicative gaps.%

The rest of this paper is organized as follows. Section \ref{systemmodel} presents the problem formulation and definitions. The computation-communication tradeoffs are presented in Section \ref{binarysection} for binary offloading and Section \ref{partialsection} for partial offloading. The conclusions are drawn in Section \ref{conclusion}.
\begin{table}\label{table}
\small
{\caption{List of important notations}
\begin{center}
\vspace{-2mm}
{\begin{tabular}{|p{5.5cm}| p{10cm}|}
\hline
\multicolumn{1}{|m{2.5cm}|}{$\mathcal{M}$ ($\mathcal{N}$)}  & \multicolumn{1}{m{10.25cm}|}{set of ENs (users)}\\ \hline
\multicolumn{1}{|m{2.5cm}|}{$h_{ij}$  ($g_{ji}$)} & \multicolumn{1}{m{10.25cm}|}{uplink (downlink) channel fading from user $j$ (EN $i$) to EN $i$ (user $j $)} \\ \hline
\multicolumn{1}{|m{2.5cm}|}{${W}_j$ ($\widetilde{W}_j$)} & \multicolumn{1}{m{10.25cm}|}{the input (output) data of the task of user $j$} \\ \hline
\multicolumn{1}{|m{2.5cm}|}{$L$ ($\widetilde{L}$)} & \multicolumn{1}{m{10.25cm}|}{the input (output) data size} \\ \hline
\multicolumn{1}{|m{2.5cm}|}{$P_{u}$, $DoF^u$  ($P_{d}$, $DoF^d$ )} & \multicolumn{1}{m{10.25cm}|}{uplink (downlink) transmission power, DoF} \\ \hline
\multicolumn{1}{|m{2.5cm}|}{$\Phi$, $|\Phi|$} & \multicolumn{1}{m{10.25cm}|}{subset of ENs, cardinality of set $\Phi$}    \\ \hline
\multicolumn{1}{|m{2.5cm}|}{$W_{j,\Phi}$ ($\widetilde{W}_{j,\Phi}$)} & \multicolumn{1}{m{10.25cm}|}{part of input (output) data of task $j$ computed by the set of ENs $\Phi$} \\ \hline
\multicolumn{1}{|m{2.5cm}|}{$\mathbf{H}$, $\bar{\mathbf{H}}_{ij}$ ($\mathbf{G}$, $\bar{\mathbf{G}}_{ji}$)} & \multicolumn{1}{m{10.25cm}|}{the uplink (downlink) channel fading matrix, symbol-extended channel from user $j$ (EN $i$) to EN $i$ (user $j$)} \\ \hline
\multicolumn{1}{|m{2.5cm}|}{$X_j(t)$, $\mathbf{X}_j$ ($Y_i(t)$, $\mathbf{Y}_i$)} & \multicolumn{1}{m{10.25cm}|}{transmitted (received) signal at time $t$, signal vector of user $j$ (EN $i$)} \\ \hline
\multicolumn{1}{|m{2.5cm}|}{$\tau^{u}$, $\tau^{u}_a$, $\tau^{u^*}$ ($\tau^{d}$, $\tau^{d}_a$, $\tau^{d^*}$)} & \multicolumn{1}{m{10.25cm}|}{NULT, achievable NULT, optimal NULT (NDLT, achievable NDLT, minmum NDLT)} \\ \hline
\multicolumn{1}{|m{2.5cm}|}{$x^l_j, \mathbf{v}^l_{j}, \bar{\mathbf{X}}_{j}, \bar{\mathbf{V}}_{j}$} & \multicolumn{1}{m{10.25cm}|}{signal stream $l$, beamforming vector $l$, signal vector, beamform matrix transmitted by user $j$ over the symbol-extended channels}  \\ \hline
\multicolumn{1}{|m{2.5cm}|}{$\bar{\mathbf{Y}}_{i}, \bar{\mathbf{Z}}_{i}$} &\multicolumn{1}{m{10.25cm}|}{signal vector, noise vector received at EN $i$ over symbol-extended channels.}  \\ \hline
\multicolumn{1}{|m{2.5cm}|}{$a_i$, $\gamma_{j,i}$} & \multicolumn{1}{m{10.25cm}|}{number of tasks assigned to EN $i$, split ratio of task $j$ assigned to EN $i$} \\ \hline
\end{tabular}}
\end{center}}
\vspace{-11mm}
\end{table}

\textbf{Notations:} $\mathbb{C}$ denotes the set of complex numbers. $\mathbb{Z}^+$ denotes the set of positive integers. $(\cdot)^{T}$ denotes the transpose. $\mathcal{K}$ denotes the set of indexes $\{1,2,\cdots\!,K\}$. $\lfloor r\rfloor$ denotes the largest
integer no greater than $r$ while $\lceil r\rceil$ denotes the minimum integer no smaller than $r$. $[a]$ denotes the set of integers $\{1,2,\cdots\!,a\}$. $[a\!:\!b]$ denotes the set of integers $\{a,a\!+\!1,\cdots\!,b\}$. $(x^j)_{j=1}^{L}$ denotes the vector $(x_1,x_2,\cdots\!,x_L)^{T}$. $\{U_j\!:\!j\in[a]\}$ denotes the set $\{U_1, U_2,\cdots\!,U_a\}$. The important notations used in this paper are summarized in Table \ref{table}.
\vspace{-2mm}
\section{Problem Formulation} \label{systemmodel}
\subsection{MEC Network Model}
\begin{figure}[t]
\centering
\includegraphics[width=2.2in, height=1.75in]{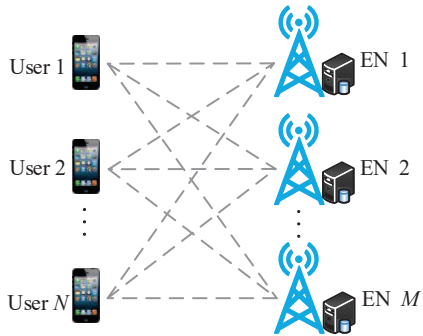}
\vspace{-3mm}
\caption{A multi-user multi-server MEC network.}
\vspace{-6mm}
\label{sysModel}
\end{figure}

We consider an infrastructure-based network consisting of $M$ single-antenna ENs (i.e. APs and BSs) which communicate with $N$ single-antenna users via a shared wireless channel, as shown in Fig. \ref{sysModel}. Each EN is equipped with computing capability and hence can serve as edge servers. The mobile users, on the other hand, do not have computing capability. They offload their computing tasks to the ENs through the uplink channel (from users to BSs) and then download the computed results back through the downlink channel (from BSs to users). The MEC platform at all ENs is implemented by container-based lightweight virtualization technologies\cite{taleb2017multi}. To simplify the analysis, we assume all ENs have the same computing capability. 
Denote by $\mathcal{M}=\{1,2,\ldots,M\}$ the set of ENs and $\mathcal{N}=\{1,2,\ldots,N\}$ the set of users. The communication links between each EN and each user are assumed to have the same path loss but experience independently distributed small-scale fading. Let $h_{ij} (g_{ji})$ denote the uplink (downlink) channel fading from user $j\in \mathcal{N}$ (EN $i\in \mathcal{M}$) to EN $i\in \mathcal{M}$ (user $j \in \mathcal{N}$). They are independent and identically distributed (i.i.d.) as some continuous distribution. The full channel state of information (CSI) is assumed to be known at all users and ENs.

The network is time-slotted. At each time slot, each user generates an independent computation task to be offloaded to the ENs for execution. On each EN, computation tasks from different users run in different processes or containers\cite{morabito2018consolidate}, with latter providing better isolation between these tasks. Modern Operating Systems are capable of running these tasks concurrently through either time-sharing or hyper-threading on multi-core processors. The computation task from each user $j$, for $j\in \mathcal{N}$, is characterized by the input data to be computed, denoted as $W_j$, with size $|W_j| = L$ bits\footnote{Note that for many applications, e.g., data compression, image processing, and video rendering, the task computing time depends on the input data size, so the input data size is an essential parameter to characterize the workload size of a task\cite{Miettinen2010Energy}.}, the computed output data, denoted as $\widetilde{W}_j$, with size $|\widetilde{W}_j| = \widetilde{L}$ bits\footnote{Here the equal size for both input data and output data from all tasks is assumed for analytical tractability. In the general case when each task has a distinct input or output data size, the problem may not be tractable.}. For simplicity, we assume the tasks are CPU-bounded and tasks with the same workload size consume an equal total CPU time whether it runs in one EN or is distributed to multiple ENs. We consider both binary and partial computation task offloading models. Binary offloading requires a task to be executed as a whole. Partial offloading, on the other hand, allows a task to be partitioned into multiple modules and executed in different nodes in a distributed manner. While binary offloading is suitable for simple tasks that are tightly integrated in structure and are not separable, partial offloading is more suitable for data-oriented applications that can be separated into multiple parts and executed in parallel, such as image processing, voice-to-text conversion, and
components rendering in $360^\circ$ VR video display. For partial offloading, it is further assumed that the task partition is exclusive without intra-task or inter-task coding and the computed output size of each subtask is proportional to its corresponding input data size.
\vspace{-2mm}
\subsection{Task Offloading Procedure}
Before the task offloading procedure begins, the system needs to decide which EN or which set of ENs should each task (for binary offloading) or subtask (for partial offloading) be assigned to for execution. We denote by $W_{j,\Phi}$ the part of task $j$ that is assigned exclusively to an arbitrary subset of ENs $\Phi\!\subseteq \!\mathcal{M}$ for computation with repetition order $|\Phi|$. For example, in partial offloading,  $W_{1,\{1,2\}}$  and $W_{1,\{2,3\}}$ denote the two disjoint subtasks of task $1$, one assigned to EN subset $\{1,2\}$ for processing, and the other assigned to EN subset $\{2,3\}$ for processing; in binary offloading, $W_{1,\{1,2,3\}}\!=\!W_1$ indicates that task 1 is assigned to EN set $\{1,2,3\}$ with all the rest $W_{1, \Phi}\!=\!\varnothing$, $\forall\Phi\!\neq\!\{1,2,3\}$. Every task must be computed. Thus, for each $j\!\in\!\mathcal{N}$, the task assignment policy must satisfy $\bigcup\limits_{\Phi\subseteq \mathcal{M}}\!W_{j,\Phi}\!=\!W_j$, and $W_{j,\Phi}\bigcap W_{j,\Psi}\!=\!\varnothing$, $\forall \Psi\!\neq\!\Phi$. By such task assignment, the set of tasks or subtasks to be computed at each EN $i \in \mathcal{M}$ is given by $\left\{W_{j,\Phi}\!: \forall \Phi \!\supseteq \!\{i\}, \forall j\!\in\! \mathcal{N}\right\}$.
\begin{definition}\label{defenition1}
For a given task assignment scheme $\{W_{j,\Phi}\}$, the computation load $r$, $1\le r\le M$, is defined as the total number of
task input bits computed at all the $M$ ENs, normalized by
the total number of task input bits from all the $N$ users, i.e.,
\begin{small}
\begin{equation}
r\triangleq \frac{\sum\limits_{i\in\mathcal{M}}\sum\limits_{j\in\mathcal{N}}\sum\limits_{\Phi\supseteq \{i\},\Phi\subseteq\mathcal{M}}|W_{j,\Phi}|}{NL}.
\end{equation}
\end{small}
\end{definition}
Similar to \cite{fundamental_tradeoff}, the computation
load $r$ can be interpreted as the average number of ENs to compute each task (for binary offloading) or each input bit (for partial offloading) and hence is a measure of computation repetition.

Given a feasible task assignment strategy $\{W_{j,\Phi}\}$, the overall offloading procedure contains two communication phases, an input data uploading phase and an output data downloading phase.
\subsubsection{Uploading phase}
Each user $j$ employs an encoding function to map its task inputs $W_j$ and channel coefficients $\mathbf{H}\!\triangleq\![h_{ij}]_{i\in\mathcal{M},j\in\mathcal{N}}$ to a length-$T^u$ codeword $\mathbf{X}_j\!\triangleq\!\left(X_{j}(t)\right)^{T^{u}}_{t=1}$, where $X_{j}(t)\!\in\!\mathbb{C}$ is the transmitted symbol at time $t\!\in\![T^{u}]$. Each codeword has an average power constraint of $P_u$, i.e., $\frac{1}{T^{u}}||\mathbf{X}_j||\!\le\! P_u$. Then, the received signal $Y_i(t) \!\in\! \mathbb{C}$ of each EN $i$ at time $t\!\in\![T^{u}]$ is given by
\begin{small}
\begin{equation}
Y_i(t)=\sum_{j \in \mathcal{N}}h_{ij}(t)X_{j}(t)+Z_{i}(t),~~~~\forall i\in\mathcal{M},
\end{equation}
\end{small}
where $Z_{i}(t)\sim\mathcal{CN}(0,1)$ is the noise at EN $i$. Each EN $i$ uses a decoding function to map received signals $\left(Y_i(t)\right)^{T^{u}}_{t=1}$ and channel coefficients $\mathbf{H}$ to the estimate $\{\hat{ W}_{j,\Phi}\!:\forall j\in\mathcal{N},\forall \Phi \supseteq \{i\}\}$
of its assigned task inputs $\left\{W_{j,\Phi}\!:\forall j\in\mathcal{N},\forall\Phi\supseteq \{i\}\right\}$. The error probability is given by
\begin{small}
\begin{equation}
P^u_e =  \max\limits_{i\in \mathcal{M}}~\mathbb{P}\Bigg( \bigcup\limits_{j\in\mathcal{N},\Phi \supseteq \{i\} }\left\{\hat{ W}_{j,\Phi} \ne W_{j,\Phi}\right\}\Bigg).
\end{equation}
\end{small}
\vspace{-3mm}
\subsubsection{Downloading phase} After receiving the assigned task input data and executing them at the server, each EN $i$ obtains the output data of its assigned tasks, $\left\{\widetilde{W}_{j,\Phi}:\forall j\in\mathcal{N},\forall \Phi \supseteq \{i\}\right\}$, and begins to transmit these computed results back to users. The computed results downloading is similar to the task uploading operation. Briefly, each EN maps the task outputs $\left\{\widetilde{W}_{j,\Phi}:\forall j\in\mathcal{N},\forall \Phi \supseteq \{i\}\right\}$ and channel coefficients $\mathbf{G}\triangleq[g_{ji}]_{j\in\mathcal{N},i\in\mathcal{M}}$ into a codeword of block length $T^d$ over the downlink interference channel, with an average power constraint of $P_d$. Each user $j$ decodes its desired task output data $\widetilde{W}_{j}$ from its received signals and obtains the estimate $\hat{\widetilde{W}}_{j}$. The error probability is given by
\begin{small}
\begin{equation}
P^d_e =  \max\limits_{j\in \mathcal{N}}~\mathbb{P}\Big( \hat{\widetilde{W}}_{j} \ne\widetilde{W}_{j}\Big).
\end{equation}
\end{small}
A task offloading policy with computation load $r$, denoted as $\left(\{W_{j,\Phi}\},\big(L, \widetilde{L}\big),r\right)$, consists of a sequence of task assignment schemes $\{W_{j,\Phi}\}$, task input uploading schemes with time $T^u$, and task output downloading schemes with time $T^d$, indexed by the task input and output data size pair $\big(L, \widetilde{L}\big)$. It is said to be feasible when the error probabilities $P^u_e$ and $P^d_e$ approach to zero when $L \to \infty$ and $\widetilde{L} \to \infty$.
\vspace{-2mm}
\subsection{Performance Metric}
\vspace{-1mm}
We characterize the performance of the considered MEC network by the computation load $r$ as well as the asymptotic communication time for task input uploading and output downloading.
\begin{definition}\label{defenition2}
The normalized uploading time (NULT) and normalized downloading time (NDLT) for a given feasible task offloading policy with computation load $r$ are defined, respectively, as
\begin{small}
\begin{align}
\tau^{u}(r) &\triangleq \lim_{P_u\to\infty} \lim_{L\to\infty} \frac{\mathbb{E}_{\mathbf{H}}[T^u]}{ L/\log P_u} \label{latency1},\\
\tau^{d}(r) &\triangleq \lim_{P_d\to\infty} \lim_{\widetilde{L}\to\infty} \frac{\mathbb{E}_{\mathbf{G}}[T^d]}{\widetilde{L}/\log P_d}.
\end{align}
\end{small}
\end{definition}
Further, the minimum NULT and NDLT are defined, respectively, as
\begin{align}
\tau^{u^*}(r) \triangleq \inf\{&\tau^{u}(r): \forall\tau^{u}(r)~\text{is achievable at the computation load}~r\},\\
\tau^{d^*}(r) \triangleq \inf\{&\tau^{d}(r): \forall\tau^{d}(r)~\text{is achievable at the computation load}~r\}.
\end{align}
The NULT (or NDLT) is similar to the normalized delivery time defined in \cite{sengupta2017fog,FanXu} for cache-aided networks, and the communication load defined in \cite{8007057} for MEC systems and \cite{li2019wireless} for wireless MapReduce systems. Note that $L/\log P_d$ (or $\widetilde{L}/\log P_d$) is the reference time to transmit the input (or output) data of $L$ (or $\widetilde{L}$) bits for one task in a Gaussian point-to-point
baseline system in the high SNR regime. Thus, an NULT (or NDLT) of $\tau^{u^*}(r)$ $\big($or $\tau^{d^*}(r)\big)$ indicates that the time required to upload (or download) the tasks of all users is $\tau^{u^*}(r)$ $\big($or $\tau^{d^*}(r)\big)$ times of this reference time period. Under such definition, the specific communication distance between users and ENs does not affect analysis, and we only assume that the channels between all users and all ENs are i.i.d..
\begin{definition}
A communication latency pair $(\tau^{u}(r), \tau^{d}(r))$ at a computation load $r$ is said to be achievable if there exists a feasible task offloading policy $\left(\{W_{j,\Phi}\},\big(L, \widetilde{L}\big),r\right)$. The optimal communication latency region is the closure of the set of all achievable communication latency pairs $\left\{(\tau^u(r), \tau^d(r)\right\}$ at all possible computation load $r$'s, i.e.,
\begin{align}
\mathscr{T}\!\triangleq\!\text{closure}\big\{&\!\!\left(\tau^{u}(r), \tau^{d}(r)\right): \forall\left(\tau^{u}(r), \tau^{d}(r)\right)~\text{is achievable},  
\forall r\in [1, M]\big\}.
\end{align}
\end{definition}
Our goal is to characterize the optimal communication latency pair at any given computation load $r$ for both binary offloading and partial offloading.
\section{Communication Latency Analysis for Binary Offloading} \label{binarysection}
In this section, we present the analysis of the optimal communication latency  pair at any given computation load, including both achievable scheme and converse, for binary offloading.
\subsection{Main Results}
\begin{figure*}
\vspace{-5mm}
\centering
\subfigure[$M=N=3$.]
{\label{innerouter1} 
 \includegraphics[width=2.8in, height=2.2in]{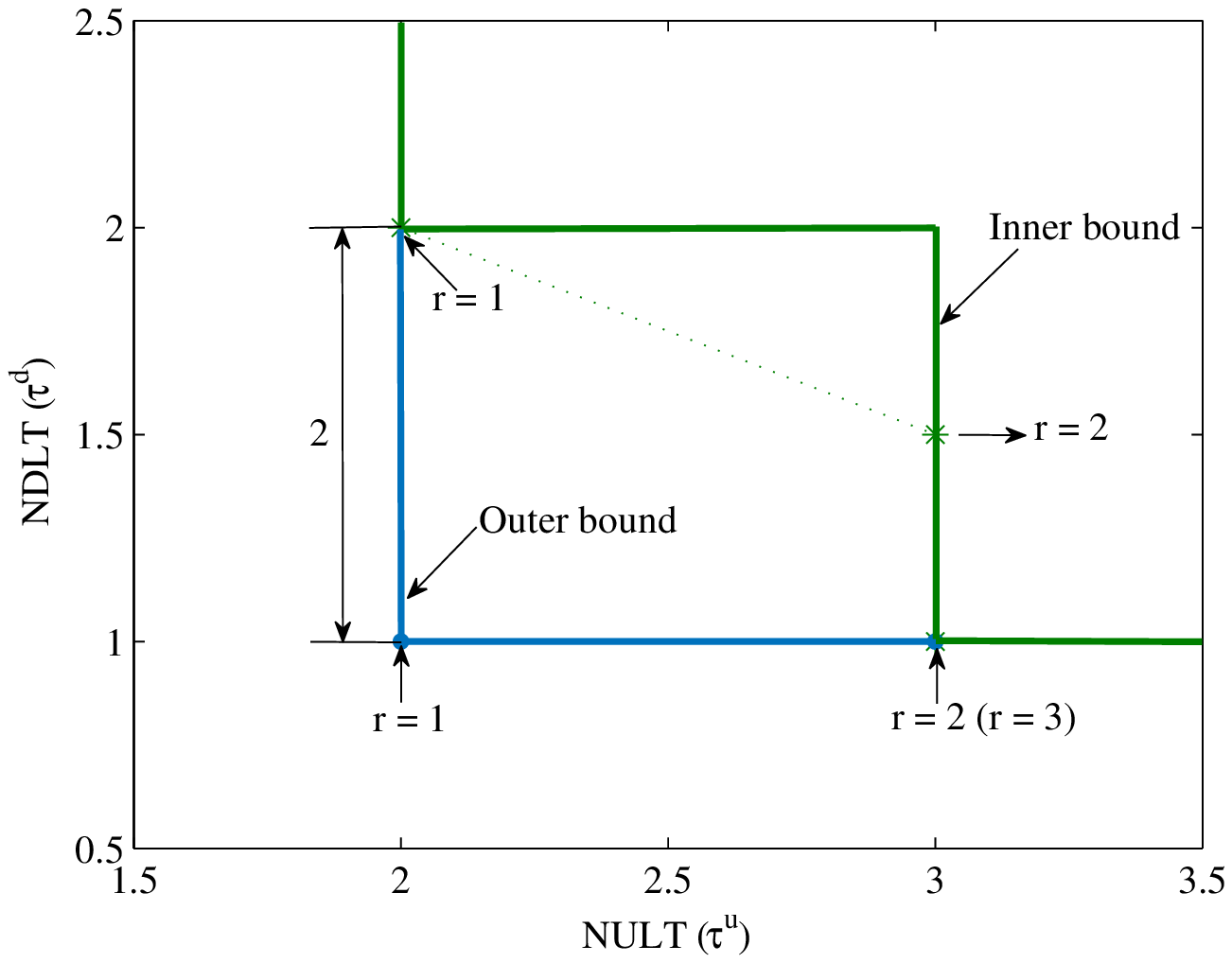}}
\hspace{12mm}
\subfigure[$M=N=10$.]
{ \label{outerbound1} 
 \includegraphics[width=2.8in, height=2.2in]{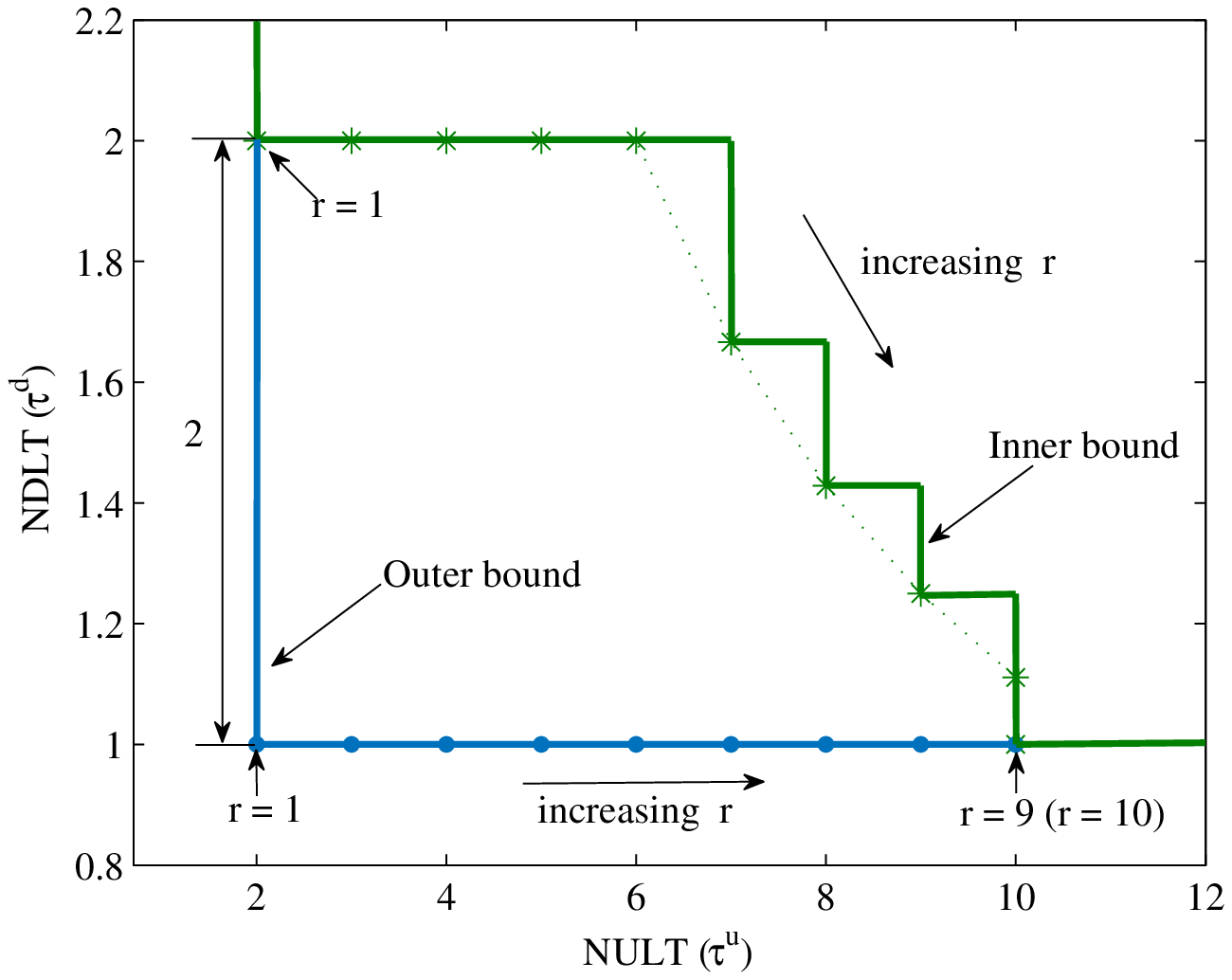}}
 \vspace{-2mm}
 \caption{The inner bound and outer bound of the optimal communication latency region for binary offloading. At a given integer computation load $r\in[M]$, the achievable NULT is optimal while the gap of NDLT is within $2$.}
\label{tradeoff_112} 
\vspace{-6mm}
\end{figure*}
\begin{theorem} \label{binary_latency region}
(Achievable result). An achievable communication latency pair $\left(\tau^{u}_a(r),\tau^{d}_a(r)\right)$ at an integer computation load $r\!\in\![M]$, for binary task offloading in the MEC network with $M$ ENs and $N$ users, is given by
\begin{small}
\begin{align}
\tau^{u}_a(r)&=\min\bigg\{1+\frac{Nr}{M},N\bigg\},\label{tau_u}\\
\tau^{d}_a(r)&=\min\bigg\{1+\frac{N}{M},\frac{N}{r}\bigg\}, \label{tau_d}
\end{align}
\end{small}
\!\!when $\frac{N}{M}\!\in\!\mathbb{Z}^+$. If $\frac{N}{M}$ is not an integer, one can always find two integers $\delta_1$ and $\delta_2$ so that $\frac{N+\delta_1}{M - \delta_2}$ is the closest integer to $\frac{N}{M}$ and the above results still hold by adding $\delta_1$ more users and deactivating $\delta_2$ ENs.
\end{theorem}
We prove the achievability of Theorem \ref{binary_latency region} in Section \ref{binary_scheme}.
\begin{theorem}\label{theorem2}
(Converse). The optimal communication latency pair $\left(\tau^{u^*}(r),\tau^{d^*}(r)\right)$ at any given computation load $r\!\in\!\left\{r\!:\!\sum_{i\in\mathcal{M}}\!a_i \!=\!Nr,a_i\!\in\![0\!:\!N],\forall i\!\in\!\mathcal{M}\right\}$, for binary task offloading in the MEC network with $M$ ENs and $N$ users, is lower bounded by
\begin{small}
\begin{align}
\tau^{u^*}(r)&\ge\min\bigg\{1+\frac{Nr}{M},N\bigg\},\\
\tau^{d^*}(r)&\ge \frac{N}{\min\{M,N\}}.
\end{align}
\end{small}
\end{theorem}
Based on Theorem \ref{binary_latency region} and Theorem \ref{theorem2}, we can obtain an inner bound denoted as $\mathscr{T}_{in}$ and an outer bound denoted as $\mathscr{T}_{out}$, respectively, of the optimal communication latency region by collecting the latency pairs $(\tau^u(r), \tau^d(r))$ at all the considered computation loads $r$'s. Fig. \ref{tradeoff_112} shows the bounds in the MEC networks with $M\!=\!N\!\in\!\{3,10\}$. Our achievable communication latency region totally contains that of \cite{8007057} that considers the repetition order for each task is $M$.
\begin{corollary} \label{corollary11}
(Optimality). At an integer computation load $r\!\in\!\![M]$, the achievable NULT in (\ref{tau_u}) is \textbf{optimal}, and the achievable NDLT in (\ref{tau_d}) is within a multiplicative gap of $2$ to its minimum.
\end{corollary}
The proofs for Theorem \ref{theorem2} and Corollary \ref{corollary11} are given in Section \ref{binaryconverse}.

Now, we demonstrate how the computation load $r$ affects the achievable communication latency $(\tau^u_a(r),\tau^d_a(r))$. By discussing the $\min$ function terms in (\ref{tau_u}) and (\ref{tau_d}), we have the monotonicity of the achievable computation-communication function  $(\tau^u_a(r),\tau^d_a(r))$:
\begin{itemize}
\item The NULT $\tau^{u}_a(r)$ increases strictly with the computation load $r$ for $1\!\le\! r\!\le\! M\!-\!\frac{M}{N}$, and then keeps a constant $N$ for $M\!-\!\frac{M}{N}\!\le\! r \!\le\! M$.
\item The NDLT $\tau^{d}_a(r)$ keeps a constant $1\!+\!\frac{N}{M}$ for $1\!\le\!r\!\le\!\frac{MN}{M+N}$, and then is \textbf{inversely proportional} to the computation load $r$ for $\frac{MN}{M+N}\!\le\!r\!\le\!M$.
\end{itemize}

\begin{remark}
The achievable computation-communication function $(\tau^u_a(r),\tau^d_a(r))$ has two corner points $\left(N,\frac{N^2}{MN\!-M}\right)$ and $\left(1\!+\!\frac{N^2}{M\!+N},1\!+\!\frac{N}{M}\right)$, corresponding to $r\!=\!M\!-\!\frac{M}{N}$ and $r\!=\!\frac{MN}{M\!+N}$, respectively. They are explained as follows:
\begin{itemize}
\item For input data uploading, before $r$ increases to $M\!-\frac{M}{N}$, the NULT is increasing since more traffic is introduced in the uplink. When $r$ grows to more than $M\!-\frac{M}{N}$, there is no need to increase the NULT since all tasks can be uploaded within $N$ time slots by using TDMA.
\item For output data downloading, before $r$ increases to $\frac{MN}{M\!+N}$, the potential transmission cooperation gain brought by computation replication cannot exceed the existing interference alignment gain without computation replication and thus the NDLT keeps fixed. When $r$ grows to more than $\frac{MN}{M\!+\!N}$, interference neutralization can be exploited, which outperforms interference alignment, and thus the NDLT begins to decrease with $r$.
\end{itemize}
\end{remark}
It can be easily proved that $M\!-\!\frac{M}{N}\!\ge\!\frac{MN}{M+N}$ for all $M,N\!\ge\!2$. Hence, we have the following remark to characterize the envelope of the inner bound of the optimal communication latency region, present the tradeoff between computation load and communication latency, and illustrate the interaction between the NULT and NDLT.
\begin{remark}
The envelope of the inner bound  $\mathscr{T}_{in}$ of the optimal communication latency region for binary offloading can be divided into three sections, each corresponding to a distinct interval of the computation load $r$:
\begin{enumerate}
\item Constant-NDLT section: $\tau^u (r)\!=\!1+\frac{Nr}{M}$, $\tau^d(r)\!=\!1+\frac{N}{M}$, when $1\!\le\! r\!\le\!\frac{MN}{M+N}$;
\item NULT-NDLT tradeoff section: $\tau^u(r)\!=\!1+\frac{Nr}{M}$, $\tau^d(r)\!=\!\frac{N}{r}$, when $\frac{MN}{M+N}\!\le\! r \!\le\! M\!-\!\frac{M}{N}$;
\item Constant-NULT section: $\tau^u(r)\!=\!N$, $\tau^d(r)\!=\!\frac{N}{r}$, when $M\!-\!\frac{M}{N}\!\le\! r \le M$.
\end{enumerate}
In particular, in the NULT-NDLT tradeoff section, as the computation load increases, the NDLT decreases in an inversely proportional way, at the expense of increasing the NULT linearly.
\end{remark}
It is seen from Fig. \ref{outerbound1} that the envelope of the inner bound is composed of three sections corresponding to three different intervals of the computation load, and the middle section at $5\!\le\!r\!\le\!9$ (dotted line) presents the NULT-NDLT tradeoff, in an inversely proportional form.
\vspace{-4.5mm}
\subsection{Achievable task offloading scheme for Theorem \ref{binary_latency region}} \label{binary_scheme}
\vspace{-0.5mm}
\subsubsection{Task assignment and uploading} \label{binary_schemeuploading}
Consider that the system parameters $M$ and $N$ satisfy $\frac{N}{M}\!\in\!\mathbb{Z}^+$ such that $Nr\!=\!Mn$ holds for $r\!\in\![M]$, where $r$ is the given integer computation load and $n$ is an integer in $[N]$. In the proposed task assignment method, we let each task be executed at exactly $r$ different ENs and let each EN execute $n$ distinct tasks with even load. Note that if $\frac{N}{M}$ is not an integer, we can inject $\delta_1$ ($\ge\!0$) tasks and let $\delta_2$ ($\ge\!0$) ENs being idle and use the remaining ENs for task offloading, such that $\frac{N+\delta_1}{M-\delta_2}$ is the integer closest to $\frac{N}{M}$, denoted as $n_1$. In this way, we still have $(N\!+\!\delta_1)r \!=\! (M\!-\!\delta_2)rn_1$, $\forall r\!\in\![M\!-\!\delta_2]$, and can use the new $N\!+\!\delta_1$ and $M\!-\!\delta_2$ to replace $N$ and $M$ to obtain the corresponding analytical results.

\begin{figure}[t]
\centering
\includegraphics[width=2.7in, height=1.6in]{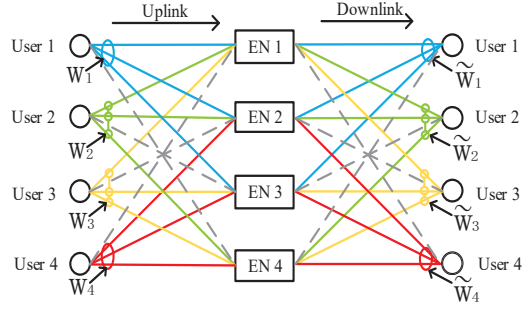}
\vspace{-3mm}
\caption{Illustration of even task assignment for $M\!=\!N\!=\!4$ and $r\!=\!3$. The tasks \emph{circularly} assigned to 4 ENs are $\{W_1,W_2,W_3\}$, $\{W_4,W_1,W_2\}$, $\{W_3,W_4,W_1\}$, and $\{W_2,W_3,W_4\}$. The uplink channel is the circular interference-multicast channel whose per-receiver DoF is $\frac{3}{4}$ achieved via TDMA, while the downlink is the circular cooperative interference channel whose per-receiver DoF is $\frac{3}{4}$ achieved by interference neutralization. Solid circle denotes that the channels inside it carry the same information.}
\vspace{-9mm}
\label{offloadModel2}
\end{figure}

To ensure even task assignment on each EN, we perform circular assignment. Specifically, the set of tasks assigned to EN $i\in\mathcal{M}$ is given by
\begin{small}
\begin{equation}
\mathcal{T}_{i}=\big\{W_{j+1}\!: j\in \left[(i\!-\!1)n: (in\!-\!1)\right]\!\!\!\!\!\!\pmod{\!N}\big\}. \label{ntask}
\end{equation}
\end{small}
\!\!An example of the task uploading for $M\!=\!N\!=\!4$ and $r\!=\!3$ is shown in Fig. \ref{offloadModel2}.

Given the above task assignment in (\ref{ntask}), the uplink channel formed by uploading the $N$ tasks to their corresponding ENs is referred to as \emph{the circular interference-multicast channel with multicast group size $r$}. This channel is different from the X-multicast channel with multicast group size $r$ defined in \cite{FanXu,DOfNiesen}, where any subset of $r$ receivers can form a multicast group, resulting in $\binom{M}{r}$ multicast groups, and each transmitter needs to communicate with all the $\binom{M}{r}$ multicast groups. In our considered circular interference-multicast channel, there are only $N$ multicast groups which are performed circularly by the $M$ receivers and each transmitter only needs to communicate with one multicast group. The optimal per-receiver DoF of this uplink channel is given as follows.
\begin{lemma}\label{lemma4}
The optimal per-receiver DoF of the circular interference-multicast channel with $N$ transmitters and $M$ receivers satisfying $\frac{N}{M}\in\mathbb{Z}^+$ and multicast group size $r$ is given by
\begin{small}
\begin{equation}
DoF^{u}_r= \max\bigg\{\frac{Nr}{Nr+M},\frac{r}{M}\bigg\},~r\in[M].          \label{dof_u}
\end{equation}
\end{small}
\end{lemma}
\emph{Proof:}
We use partial interference alignment to achieve a DoF of $\frac{n}{n\!+\!1}\!=\!\frac{Nr}{Nr\!+\!M}$ for each receiver, it is similar to the achievable scheme of the DoF of $K$-user interference channels in \cite{K_dof}. We can also achieve a DoF of $\frac{r}{M}$ by using TDMA. The maximum achievable DoF is thus given by (\ref{dof_u}). The detailed achievable scheme and optimality proof are given in Appendix \ref{lemma4proof}.

The per-receiver rate of this channel in the high SNR regime can be
approximated as $DoF^{u}_{r}\!\times\!\log\!P_u\!+\!o(\log\!P_u)$. The traffic load for each EN to receive its assigned tasks is $\frac{Nr}{M}L$ bits, then the uploading time can be approximately given by $T^u\!=\!\frac{\frac{Nr}{M}L}{DoF^{u}_{r}\log\!P_u\!+o(\log\!P_u)}$. Let $P_u\!\to\!\infty$ and $L\!\to\!\infty$, by Definition \ref{defenition2}, the NULT for each EN at computation load $r$ is given by
\begin{small}\begin{equation} \label{T^{u'}(m)}
\tau^{u}_a(r)=\frac{\frac{Nr}{M}}{DoF^{u}_{r}}=\min\bigg\{\frac{Nr}{M}+1,N\bigg\},~~r\in[M].
\end{equation} \end{small}
\vspace{-3mm}
\subsubsection{Results downloading}
After computing all offloaded tasks, ENs begin to transmit the computed results back to users via downlink channels. Recall that each task is computed at $r$ different ENs and each EN $i$ has the computed results of $n$ different tasks  $\mathcal{T}_{i}$, as given in (\ref{ntask}). Each user $j$ wants the computed results $\widetilde{W}_{j}$, $\forall j\in\mathcal{N}$, which is owned in $r$ different ENs. Multiple ENs with the same computed results can exploit transmission cooperation to neutralize interferences across users \cite{coBC,5752451}. The computation results downloading for $M=N=4$ and $r=3$ is shown in Fig. \ref{offloadModel2}. We refer to the downlink channel formed by downloading the $N$ tasks as \emph{the circular cooperative interference channel with transmitter cooperation group size $r$}. This channel is different from the cooperative X channel with transmitter cooperation group size $r$ defined in \cite{FanXu,TxCache}, where any subset of $r$ transmitters can form a cooperation group, resulting in $\binom{M}{r}$ groups in total, and each transmitter cooperation group has messages to send to all receivers. In our considered downlink channel, there are only $N$ cooperation groups which are performed circularly by the $M$ transmitters and each group only needs to communicate with one receiver. An achievable per-receiver DoF of this downlink channel is given as below.
\begin{lemma}\label{lemma5}
An achievable per-receiver DoF of the circular cooperative interference channel with $M$ transmitters and $N$ receivers satisfying $\frac{N}{M}\!\in\!\mathbb{Z}^+$ and transmitter cooperation group size $r$ is given by
\begin{small}
\begin{equation}\label{dofdp}
DoF^{d}_{r}= \max\bigg\{\frac{M}{N+M},\frac{r}{N}\bigg\}, ~~r\in[M],
\end{equation}
\end{small}
and it is within a multiplicative gap of 2 to the optimal DoF.
\end{lemma}
\emph{Proof:}
When $r\!=\!1$, we use partial interference alignment scheme to achieve a DoF of $\frac{M}{N+M}$ for each receiver. The achievable scheme is similar to that for the $M\!\times \!N$ user X channel \cite{Xdof}. We then compare it to a DoF of $\frac{1}{N}$ achieved via TDMA. When $r\!\ge\!2$, we prove that the achievable per-receiver DoF is $\max\!\left\{\!\frac{M}{N+M},\frac{r}{N}\!\right\}$, where we first use interference neutralization to achieve a DoF of $\frac{r}{N}$ for each receiver, and then compare it with the per-receiver DoF of $\frac{NM}{N+M}$ achieved by only using interference alignment. Summarizing these two cases, we have (\ref{dofdp}). Please refer to Appendix \ref{lemma5proof} for the detailed achievable scheme and optimality proof.

The per-receiver channel rate in the high SNR regime can be
approximated as $DoF^{d}_{r}\!\times\!\log\!P_d\!+\!o(\log\!P_d)$. The traffic load for each user to download its task output data is $\widetilde{L}$ bits, then the downloading time can be approximately given by $T^d\!=\!\frac{\widetilde{L}}{DoF^{d}_{r}\log\!P_d+o(\log\!P_d)}$. Let $P_d\!\to\!\infty$ and $\widetilde{L}\!\to\!\infty$, by Definition \ref{defenition2}, the NDLT for each user at computation load $r$ is given by \begin{small}
\begin{equation} \label{T^{d'}_m}
\tau^{d}_a(r) =\!\frac{1}{DoF^{d}_{r} }  =
 \min\!\bigg\{\frac{N}{M}\!+\!1,\frac{N}{r}\!\bigg\}, ~~r\!\in[M].\\
\end{equation} \end{small}
By (\ref{T^{u'}(m)}) and (\ref{T^{d'}_m}), we thus have the achievable communication latency pair $(\tau^{u}_a(r),\!\tau^{d}_a(r))$ at an integer computation load $r\!\in\![M]$ for binary offloading.

\vspace{-4mm}
\subsection{Proof of Converse for Theorem \ref{theorem2} and Corollary \ref{corollary11}}\label{binaryconverse}
\vspace{-1mm}
\subsubsection{Lower bound and optimality of NULT}\label{binaryconverseup}
We prove the lower bound of the NULT at any given computation load $r\!\in\!\left\{r\!:\!\sum^{M}_{i=1}\!a_i \!=\!Nr,a_i\!\in\![0\!:\!N],\forall i\!\in\!\mathcal{M}\right\}$, i.e., $\tau^{u^*}(r)\!\ge\!\min\!\left\{\!\frac{Nr}{M}\!+\!1,N\!\right\}$. First, we use genie-aided arguments to derive a lower bound on the NULT of any given feasible task assignment policy with computation load $r$. Then, we optimize the lower bound over all feasible task assignment policies to obtain the minimum NULT for a given computation load $r$.

Given a computation load $r$. Consider an arbitrary task assignment policy where the number of tasks assigned to each EN $i$ is denoted as $a_i$, $\forall i\in\mathcal{M}$, and satisfies
\begin{small}
\begin{align}
&\sum_{i\in\mathcal{M}}a_i =Nr ,\label{const1111}\\
&~a_i\in[0:N],~~i\in\mathcal{M}. \label{const2222}
\end{align}
\end{small}
\!\!\!Note that we only need consider $a_i>0$ case since $a_i=0$ means no task is assigned to EN $i$ and we can remove EN $i$ from the EN set $\mathcal{M}$, which will not change the results. Consider the following three disjoint subsets of task input data (or message):
\begin{small}
 \begin{align}
\mathcal{W}_{r} &= \{W_{j,\mathcal{S}_j}: j\in\mathcal{N}, i\in\mathcal{S}_j\},\\
\mathcal{W}_{t} &= \{W_{j,\mathcal{S}_j}:j=t_o,i\notin\mathcal{S}_j\},\\
\overline{\mathcal{W}} &= \{W_{j,\mathcal{S}_j}: j\ne t_o~\text{and}~i\notin\mathcal{S}_j\},
\end{align}
\end{small}
\!\!\!where $W_{j,\mathcal{S}_j}$ denotes the input message of task $j$ that is assigned to all ENs in subset $\mathcal{S}_j$, and $t_o$ denotes one of the users that do not offload their tasks to EN $i$, i.e.,  $\mathcal{W}_{r}\cap\mathcal{W}_{t}\!=\!\varnothing$. It is seen that the set $\mathcal{W}_{r}$ indicates the messages that EN $i$ need decode, i.e., $|\mathcal{W}_{r}|\!=\!a_i$; The set $\mathcal{W}_{t}$ is a nonempty set with cardinality $|\mathcal{W}_{t}|\!=\!1$ when EN $i$ is not assigned all $N$ tasks (or $a_i\!<\!N$), since user $t_o$ exists in this case; Otherwise, we have $\mathcal{W}_{t}\!=\!\varnothing$ for $a_i\!=\!N$. We will show that set $\mathcal{W}_{r}\!\cup\!\mathcal{W}_{t}$ has the maximum number of messages that can be decoded by EN $i$.

Let a genie provide the messages $\overline{\mathcal{W}}$ to all ENs, and additionally provide messages $\mathcal{W}_{r}$ to ENs in $\mathcal{M}/\{i\}$.
The received signal of EN $i$ can be represented as
\begin{small}
\begin{align}
\hat{\mathbf{y}}_i &= \sum\limits^{M}_{j=1,\ne t_o}\mathbf{H}_{ij}\mathbf{x}_{j} + \mathbf{H}_{it_o}\mathbf{x}_{t_o}+\hat{\mathbf{z}}_i,
\end{align}
\end{small}
\!\!\!where $\mathbf{H}_{ij}$, $\mathbf{x}_{j}$, $\mathbf{z}_i$ are diagonal matrices representing the channel coefficients from user $j$ to EN $i$, signal transmitted by user $j$, noise received at EN $i$, over the block length $T^u$, respectively.
Note that we reduce the noise at EN $i$ from $\mathbf{z}_i$ to $\hat{\mathbf{z}}_i$ by a fixed amount such that its received signal $\mathbf{y}_i$ can be replaced by $\hat{\mathbf{y}}_i$. The ENs in $\mathcal{M}/\{i\}$ have messages $\overline{\mathcal{W}}\!\cup\!\mathcal{W}_{r}$, which do not include the message of user $t_o$. Using these genie-aided information, each EN $k\!\in\!\mathcal{M}/\{i\}$ can compute the transmitted signals $\{\mathbf{x}_j\!:j\!\ne\!t_o\}$ and subtract them from the received signal. Thus, the received signal of EN $k\!\ne\!i$ can be rewritten as
\begin{small}
\begin{equation} \label{equation1122}
\bar{\mathbf{y}}_k = \mathbf{y}_k - \sum\limits^{N}_{j=1,\ne t_o}\mathbf{H}_{kj}\mathbf{x}_{j} = \mathbf{H}_{kt_o}\mathbf{x}_{t_o} + \mathbf{z}_k.
\end{equation}
\end{small}
\!\!Since the message $\mathcal{W}_{t}$ is intended for some ENs in $\mathcal{M}/\{i\}$, denoted as $\mathcal{R}_{t}$, the ENs in $\mathcal{R}_{t}$ can decode it. By Fano{'}s inequality and (\ref{equation1122}), we have
\begin{small}
\begin{equation} \label{fano1}
H(\mathcal{W}_{t}|\mathbf{y}_k,\overline{\mathcal{W}},\mathcal{W}_{r}) \le T^u \epsilon, ~~k\in\mathcal{R}_{t}.
\end{equation}
\end{small}
\!\!Consider EN $i$, it can decode messages $\mathcal{W}_{r}$ intended for it. By Fano{'}s inequality, we have
\begin{small}
\begin{equation}\label{Fano11}
H(\mathcal{W}_{r}|\hat{\mathbf{y}}_i,\overline{\mathcal{W}}) \le |\mathcal{W}_{r}|T^u\epsilon .
\end{equation}
\end{small}
\!\!Using genie-aided messages $\overline{\mathcal{W}}$ and
decoded messages $\mathcal{W}_{r}$, EN $i$ can compute the transmitted signals $\{\mathbf{x}_j\!:j\!\ne\!t_o\}$, and subtract them from the received signal. We thus have
\begin{small}
\begin{equation}
\bar{\mathbf{y}}_i = \hat{\mathbf{y}}_i - \sum\limits^{N}_{j=1,\ne t_o}\mathbf{H}_{ij}\mathbf{x}_{j} = \mathbf{H}_{it_o}\mathbf{x}_{t_o} + \hat{\mathbf{z}}_i.
\end{equation}
\end{small}
\!\!By reducing noise and multiplying the constructed signal $\bar{\mathbf{y}}_i$ at EN $i$ by $\mathbf{H}_{kt_o}\mathbf{H}^{-1}_{it_o}$, we have
\begin{small}
\begin{equation}
\bar{\mathbf{y}}^{k}_i  = \mathbf{H}_{kt_o}\mathbf{H}^{-1}_{it_o} \bar{\mathbf{y}}_i =  \mathbf{H}_{kt_o}\mathbf{x}_{t_o} + \hat{\mathbf{z}}^{'}_k,
\end{equation}
\end{small}
\!\!where $\hat{\mathbf{z}}^{'}_k$ represents the reduced noise. It is seen that $\bar{\mathbf{y}}^{k}_i $ is a degraded version of $\bar{\mathbf{y}}_k $ at EN $k$ in $\mathcal{R}_{t}$,
so EN $i$ must be able to decode the messages that ENs in $\mathcal{R}_{t}$ can decode. Thus, we have
\begin{small}
\begin{equation}\label{fano2}
H(\mathcal{W}_{t}|\hat{\mathbf{y}}_i,\overline{\mathcal{W}},\mathcal{W}_{r}) \le H(\mathcal{W}_{t}|\mathbf{y}_k,\overline{\mathcal{W}},\mathcal{W}_{r}) \le T^u \epsilon, ~~i\in\mathcal{R}_{t}.
\end{equation}
\end{small}
\!\!All the above changes including genie-aided information, receiver cooperation, and noise reducing can only improve capacity. Therefore, we have the following chain of inequalities,
\begin{small}
\begin{align}
\!\!\!\!\!\!(|\mathcal{W}_{r}|+|\mathcal{W}_{t}|)L &= H(\mathcal{W}_{r},\mathcal{W}_{t})\\
\!\!\!\!\!\!&\stackrel{(a)}{=}H(\mathcal{W}_{r},\mathcal{W}_{t}|\overline{\mathcal{W}})\\
\!\!\!\!\!\!&\stackrel{(b)}{=}I(\mathcal{W}_{r},\mathcal{W}_{t};\hat{\mathbf{y}}_i|\overline{\mathcal{W}})\!+\! H(\mathcal{W}_{r},\mathcal{W}_{t}|\hat{\mathbf{y}}_i,\overline{\mathcal{W}})\\
\!\!\!\!\!\!&\stackrel{(c)}{=}I(\mathcal{W}_{r},\mathcal{W}_{t};\hat{\mathbf{y}}_i|\overline{\mathcal{W}})\!+\! H({\mathcal{W}_{r}|\hat{\mathbf{y}}_i,\overline{\mathcal{W}}})\!+\!H({\mathcal{W}_{t}|\hat{\mathbf{y}}_i,\mathcal{W}_{r},\overline{\mathcal{W}}})\!\!\!\\
\!\!\!\!\!\!&\stackrel{(d)}{\le} I(\mathcal{W}_{r},\mathcal{W}_{t};\hat{\mathbf{y}}_i|\overline{\mathcal{W}})\!+\! |\mathcal{W}_{r}|T^u \epsilon\!+\!T^u\epsilon\\
\!\!\!\!\!\!&\stackrel{(e)}{\le} I(\mathbf{x}_1,\mathbf{x}_2,\cdots,\mathbf{x}_{a_i},\mathbf{x}_{t_o};\hat{\mathbf{y}}_i|\overline{\mathcal{W}})\!+\! (|\mathcal{W}_{r}|+1) T^u\epsilon \\
\!\!\!\!\!\!&\stackrel{(f)}{\le}T^u \log P_u \!+\!(|\mathcal{W}_{r}|+1) T^u\epsilon,
\end{align}
\end{small}
\!\!where (a) is due to the independence of messages, (b) and (c) follow from the chain rule, (d) uses Fano{'}s inequalities (\ref{Fano11}) and (\ref{fano2}), (e) is the data processing inequality, and (f) uses the DoF bound of the MAC channel. By dividing on $\frac{L}{\log P_u}$, and taking $P_u\!\to\!\infty$ and
$\epsilon\!\to\!0$, we have
$\tau^u \ge |\mathcal{W}_{r}|+|\mathcal{W}_{t}| = \min\{a_i + 1,N\}$.

Thus, for any given feasible task assignment $\mathbf{a}\!\triangleq\![a_i]_{i\in\mathcal{M}}$, the NULT satisfies $\tau^u \!\ge\!  \min\{a_i + \!1,\!N\}$, $\forall i\!\in\!\mathcal{M}$, i.e., the minimum NULT of the task assignment policy $\mathbf{a}$ is lower bounded by
\begin{small}
\begin{equation}
\tau^{u^*}(r,\mathbf{a}) \ge \max\limits_{ i\in\mathcal{M}}\min\left\{a_i + 1, N\right\} = \min\Big\{\max\limits_{ i\in\mathcal{M}} a_i + 1, N\Big\}.
\end{equation}
\end{small}
\!\!Hence, the minimum NULT of all feasible task assignment is given by $\tau^{u^*}\!(r)\!=\!\min\limits_{\mathbf{a}}\tau^{u^*}\!(r,\mathbf{a})$. It can be lower bounded by the optimal solution of the following linear programming problem,
\begin{small}
\begin{align}
\mathrm{P1}:\quad&\min\limits_{\mathbf{a}}~\min\Big\{\max\limits_{ i\in\mathcal{M}} a_i + 1, N\Big\} \nonumber\\
&~\mathnormal{s.t.}\quad(\ref{const1111}), (\ref{const2222})\nonumber
\end{align}
\end{small}
\!\!By relaxing the integer constraint $a_i\!\in\![0\!:\!N]$ into a real-value constraint $0\!\le\!a_i\!\le\!N$, the optimal solution is still a lower bound of the minimum NULT $\tau^{u^*}(r)$. Since the objective is equivalent to minimizing the term $\max\limits_{\forall i\in\mathcal{M}}a_i$, the optimal solution can be obtained easily as $a^*_i\!=\!\frac{Nr}{M}$, $\forall i\!\in\!\mathcal{M}$. Hence, the minimum NULT is lower bounded by
\begin{small}
\begin{equation}
\tau^{u^*}(r)\ge\min\left\{\frac{Nr}{M}+ 1, N\right\}. \label{lowerbound_tauu}
\end{equation}
\end{small}
\!\!The proof of the lower bound of NULT is thus completed. Comparing (\ref{lowerbound_tauu}) with (\ref{tau_u}) in Theorem \ref{binary_latency region}, we see that they are the same. Thus, the achievable NULT in (\ref{tau_u}) is optimal.
\subsubsection{Lower bound and gap of NDLT}\label{binaryconversedown}
Let $\mathbf{x_i}$ denote the signal transmitted by each EN $i$, and $\mathbf{y_j}$ the signal received at each user $j$, over the block length $T^d$.
Consider the $N$ computed results decoded by $N$ users, we have the following chain of inequalities,
\begin{small}
\begin{align}
\!\!\!\!\!&N \widetilde{L} \!=\! H(\widetilde{W}_{1},\cdots,\widetilde{W}_{N})\nonumber\\
\!\!\!\!\!&\!=\!I(\widetilde{W}_{1},\!\cdots\!,\widetilde{W}_{N};\mathbf{y}_1,\!\cdots\!,\mathbf{y}_N)\!+\! H(\widetilde{W}_{1},\!\cdots\!,\widetilde{W}_{N}|\mathbf{y}_1,\!\cdots\!,\mathbf{y}_N)\!\!\\
\!\!\!\!\!&\!\stackrel{(g)}{\le}\!I(\widetilde{W}_{1},\cdots,\widetilde{W}_{N};\mathbf{y}_1,\cdots,\mathbf{y}_N) \!+\! \sum\limits_{j\in\mathcal{N}}H(\widetilde{W}_{j}|\mathbf{y}_j) \\
\!\!\!\!\!&\!\stackrel{(h)}{\le}\!I(\mathbf{x}_1,\mathbf{x}_2,\cdots,\mathbf{x}_M;\mathbf{y}_1,\cdots,\mathbf{y}_N)\!+\!N T^d\epsilon \\
\!\!\!\!\!&\!\stackrel{(i)}{\le}\!\min\{M,N\} T^d \log P_d \!+\!N T^d\epsilon,
\end{align}
\end{small}
\!\!\!where $(g)$ follows from $H(\widetilde{W}_{1},\cdots,\widetilde{W}_{N}|\mathbf{y}_1,\cdots,\mathbf{y}_N)\!\le\!\sum\limits_{j\in\mathcal{N}}H(\widetilde{W}_{j}|\mathbf{y}_1,\cdots,\mathbf{y}_N)\!\le\!\sum\limits_{j\in\mathcal{N}}H(\widetilde{W}_{j}|\mathbf{y}_j)$,
$(h)$ follows from the data processing inequality and Fano{'}s inequality, and $(i)$ uses the capacity bound of the MISO broadcast channel with a $M$-antenna transmitter and $N$ single-antenna receivers. By dividing on $\frac{\widetilde{L}}{\log P_d}$, and taking $P_d\!\to\!\infty$ and
$\epsilon\!\to\!0$, we have
\begin{small}
\begin{align}
\tau^d \ge \frac{N}{\min\{M,N\}}. \label{lowerbound_tadd}
\end{align}
\end{small}
\!\!Hence, the minimum NDLT is lower bounded by $\tau^{d^*}\!\ge\!\frac{N}{\min\{M,N\}}$. It can be easily proved that the multiplicative gap between the achievable NDLT $\tau^{d}_a(r)$ in Theorem \ref{binary_latency region} and this lower bound is within $2$ for $\frac{N}{M}\in\mathbb{Z}^+$, i.e.,
$\frac{\tau^{d}_a(r)}{{\tau^{d}}^{*}(r)}\!\le\!\frac{\min \left\{\frac{N}{M} + 1,\frac{N}{r} \right\}}{N/M}\!\le\! 2$. We complete the proof of the lower bound and gap of the NDLT for binary offloading.
\vspace{-4mm}
\subsection{Numerical Example}\label{simulations}
\vspace{-1mm}
We use a numerical example to show the computing times and actual transmission times for binary offloading. Consider $4$-user $4$-server MEC networks, and each user has a task (e.g. image processing) to be offloaded to ENs for processing. The task input file size before processing is $L\!=\!512$ KB, the CPU frequency of each EN is $f\!=\!4.1$ GHz (e.g. Inter Xeon E3-1286), the number of CPU cycles required to process one-byte data is $c\!=\!330$ \cite{Miettinen2010Energy}. For the uplink and downink channel, we consider the \emph{normalized} Rayleigh channel fading and \emph{normalized} noise power, the uplink channel bandwidth $B_u\!=\!$ 10 MHz, and the downlink channel bandwidth $B_d\!=\!$ 10 MHz. We use one-shot linear precoding to realize our task uploading and results downloading schemes. The simulations are averaged over 50000 independent channel realizations. We analyze the cases when computation load $r\!=\!1,3,4.$
\begin{table}
\small
{\caption{Actual computing, uploading, and downloading times (sec)} \label{tablee}
\begin{center}
\vspace{-6.5mm}
\begin{tabular}{c|ccc|ccc|ccc}
\hline
\multirow{2}{*}{ } & \multicolumn{3}{c|}{$P^u\!=\!P^d\!=\!10$ dB} & \multicolumn{3}{c|}{$P^u\!=\!P^d\!=\!20$ dB} & \multicolumn{3}{c}{$P^u\!=\!P^d\!=\!30$ dB} \\
\cline{2-10}
&  $r\!=\!1$      &  $r\!=\!3$   &   $r\!=\!4$
&  $r\!=\!1$      &  $r\!=\!3$   &   $r\!=\!4$
&  $r\!=\!1$      &  $r\!=\!3$   &   $r\!=\!4$ \\
\hline
$T^c$\!\!\!\!\!\!\!\!          &0.04                           &0.13                     & 0.17          &0.04                           &0.13                     & 0.17         &0.04                           &0.13                     & 0.17     \\
$T^u$\!\!\!\!\!\!\!\!            &0.45                          &0.69                    & 0.78                   & 0.24           & 0.32           & 0.34          & 0.17           & 0.20           & 0.21          \\
\hline
\!\!\!\!\!\!\!\!\!\!\!\!\!\!\!\!\!$\widetilde{L}\!=\!3L$: ~~$T^d$           &1.35                          & 1.79                    & 0.84                   &0.74           & 0.63            & 0.33          &0.5          & 0.32           &0.19          \\
~~~~~~~~~~Total time         &1.84                          &2.61                    &1.79                   & 1.02           &1.08         &0.84        &0.71           & 0.65         &0.57           \\
\hline
\!\!\!\!\!\!\!\!\!\!\!\!\!\!\!\!\!$\widetilde{L}\!=\!4L$: ~~$T^d$           &1.8                          & 2.38                    & 1.12                   &0.98           & 0.84            & 0.44          &0.66           & 0.43           &0.25           \\
~~~~~~~~~~Total time          &2.29                          &3.2                    &2.07                   & 1.26           &1.29          &0.95        &0.87           & 0.76         &0.63           \\
\hline
\end{tabular}
\end{center}}
\vspace{-11mm}
\end{table}

Table \ref{tablee} shows the results at different transmission power. Note that the downloading time $T^d$ is evaluated at 2 cases when $\widetilde{L}\!=\!3L$ and $\widetilde{L}\!=\!4L$. It is seen that replicating tasks on all 4 servers can always significantly reduce the end-to-end task execution time. For example, for $\widetilde{L}\!=\!4L$, when $P^u\!=\!P^d\!=\!20$ dB, the total time at $r\!=\!4$ is decreased by $24.6\%$ compared to that at $r\!=\!1$. It is also observed that replicating tasks on 3 servers may cause more total times in the low SNR region. For instance, for $\widetilde{L}\!=\!3L$, when $P^u\!=\!P^d\!=\!20$ dB, the total time at $r\!=\!3$ is increased by $5.9\%$ compared with that at $r\!=\!1$. This is because in this case, the reduction on the downloading time brought by transmission cooperation is not able to compensate the increase in the uploading time and computing time. The details of simulations are given below.

\subsubsection{Computing times}
For a given computation load $r$, the number of user tasks assigned to each server is \begin{small}$\frac{Nr}{M}$\end{small}. Each server runs these tasks concurrently (through either time-sharing or hyper-threading on multi-core processors) with computing power fully utilized. Thus, the computing time of each EN at computation load $r$ can be evaluated as  \begin{small}$T^c(r)\!=\!\frac{LNcr}{fM}$\end{small}. Note that $T^c$ increases linearly with the input data size $L$ and the computation load $r$. At $r\!=\!1,3,4$, the actual computing time is given by \begin{small}$T^c(1)\!\approx\!0.04$\end{small} sec, \begin{small}$T^c(3)\!\approx \!0.13$\end{small} sec, and \begin{small}$T^c(4)\!\approx \!0.17$\end{small} sec. 
\subsubsection{Uploading times}
Consider one-shot linear precoding, the achievable transmission schemes for all $r\!=\!1,3,4$ are the same, which is to let the users upload their tasks on orthogonal resource blocks, either in time domain using TDMA or in frequency domain using FDMA. We focus on user $1$. When $r\!=\!3$, by our task assignment scheme in (\ref{ntask}), user $1$ multicasts its task to EN 1-3. Then the transmission rate for user $1$ is given by，
\begin{small}
\begin{equation}
R^u_{3} = \mathbb{E}_{\mathbf{h}}\left[\frac{1}{4}B_u\log\left(1+\min\{|h_{11}|^2,|h_{21}|^2,|h_{31}|^2\}P^u\right)\right],
\end{equation}
\end{small}
\!\!and the actual uploading time for user $1$ is given by
\begin{small}$T^u(3)\!=\!\frac{L}{R^u_{3}}$\end{small}.
Similarly, for $r\!=\!1,4$, we can obtain the actual uploading times \begin{small}$T^u(1)$\end{small}, \begin{small}$T^u(4)$\end{small} for user $1$.
\subsubsection{Downloading times} Consider one-shot linear precoding, the achievable transmission schemes are different for different computation load $r$'s. When $r\!=\!3$, the computed results on EN 1-4 are given as \begin{small}$\{\widetilde{W}_1, \widetilde{W}_2, \widetilde{W}_3\}$, $\{\widetilde{W}_4, \widetilde{W}_1, \widetilde{W}_2\}$, $\{\widetilde{W}_3, \widetilde{W}_4, \widetilde{W}_1\}$, $\{\widetilde{W}_2, \widetilde{W}_3, \widetilde{W}_4\}$\end{small}. The achievable transmission scheme is ZF precoding. In specific, we split each results \begin{small}$\widetilde{W}_j$\end{small} into $3$ sub-results \begin{small}$\widetilde{W}_{j,k}$\end{small}, $k\!\in\![3]$, and each time $4$ ENs can cooperatively transmit $3$ different sub-results and their copies desired by $3$ users back to these $3$ users, so all sub-results can be divided into 4 groups and transmitted back on orthogonal resource blocks. By using ZF precoding, for user $1$, the precoder \begin{small}$v^i_{1,k}$\end{small} for \begin{small}$\widetilde{W}_{1,k}$\end{small} on EN $i$ are designed as
\begin{small}$
v^i_{1,k}\!=\!\sqrt{4P^d}\frac{V^i_{1,k}}{||\mathbf{V}||},i\!\in\![3],k\!\in\![3],
$\end{small}
where we consider all precoders of 4 ENs have a total power constraint $4P^d$, and
\begin{small}
$\mathbf{V}\!=\!\left(V^i_{j,k}\right)_{i\in\mathcal{N}_j,j\in[4],k\in[3]}$
\end{small} is a $1\!\times\!36$ vector, \begin{small}$V^i_{j,k}$\end{small} is the precoder for sub-results \begin{small}$\widetilde{W}_{j,k}$\end{small} on EN $i$ before it is normalized as \begin{small}$v^i_{j,k}$\end{small}, \begin{small}$\mathcal{N}_j$\end{small} is the set of ENs with results \begin{small}$\widetilde{W}_{j}$\end{small}.

We design of precoders \begin{small}$\{V^i_{j,k}\}$\end{small}. Consider the $3$ sub-results \begin{small}$\{\widetilde{W}_{j_p,k_p}\!: p\!\in\![3]\}$\end{small} and their copies desired by $3$ users are transmitted together using ZF precoding, the precoder \begin{small}$V^{i_p}_{j_p,k_p}$\end{small} for \begin{small}$\widetilde{W}_{j_p,k_p}$\end{small} on EN \begin{small}$i_p\!\in\!\mathcal{N}_{j_p}$\end{small} is given by \begin{small}$V^{i_p}_{j_p,k_p}\!=\!(-1)^{I(i_p)}\det(\mathbf{G}_{\bar{j_p},\bar{i_p}})$\end{small}, \begin{small}$\mathbf{G}_{\bar{j_p},\bar{i_p}}$\end{small} is the $2\!\times\!2$ submatrix of channel coefficient matrix \begin{small}$\mathbf{G}\!=\![g_{tq}]_{t\in\{j_1,j_2,j_3\},q\in\mathcal{N}_{j_p}}$\end{small} that does not contain the row vector \begin{small}$(g_{j_p,q})_{q\in\mathcal{N}_{j_p}}$\end{small} and column vector \begin{small}$(g_{t,i_p})_{t\in\{j_1,j_2,j_3\}}$\end{small} of \begin{small}$\mathbf{G}$\end{small}, \begin{small}$I(i_p)$\end{small} is the index order of element $i_p$ in \begin{small}$\mathcal{N}_{j_p}$\end{small}.
So the downlink transmission rate of \begin{small}$\widetilde{W}_{1,k}$\end{small} is given by
\begin{small}\begin{align}
R^d_{1,k} &= \mathbb{E}_{\mathbf{g}}\Bigg[\frac{1}{4}B_d\log\bigg(1+\frac{4P^d\big|\sum\limits^3_{i=1}g_{1i}(-1)^{1+k}V^i_{1,k}\big|^2}{||\mathbf{V}||^2}\bigg)\Bigg], ~k\!\in\![3],
\end{align} \end{small}
\!\!and the actual downloading time of \begin{small}$\widetilde{W}_1\!=\!\{\widetilde{W}_{1,k}\!:k\!\in\![3]\}$\end{small} is given by
\begin{small}$
T^d(3) \!=\!\max\limits_{k\in[3]}\frac{\widetilde{L}/3}{R^d_{1,k}}
$\end{small}.
For $r\!=\!1,4$, we can also obtain the actual downloading times \begin{small}$T^d(1)$\end{small} and \begin{small}$T^d(4)$\end{small} for user 1. Due to the page limit, the simulation details of \begin{small}$T^d(1)$\end{small} and \begin{small}$T^d(4)$\end{small} are omitted, here.

Summing the above computing time, uploading time, and downloading time, we can get the total execution time for task offloading, as shown in Table \ref{tablee}.
\vspace{-3mm}
\section{Communication Latency Analysis for Partial Offloading} \label{partialsection}
\vspace{-1mm}
In this section, we present the analysis of the optimal communication latency pair at any given computation load, including achievable scheme and converse, for partial offloading.
\vspace{-4.5mm}
\subsection{Main Results}
\vspace{-0.5mm}
\begin{theorem} \label{theorem3}
(Achievable result). An achievable communication latency pair $(\tau^{u}_a(r),\tau^{d}_a(r))$ at an integer computation load $r\!\in\![M]$, for partial task offloading in the MEC network with $M$ ENs and $N$ users, is given by
\begin{small}
\begin{align}
\tau^{u}_a(r)&=\frac{N-1}{M}r+1, \label{tau_u_p2} \\
\tau^{d}_a(r)&=\max\left\{\frac{N-r}{M}+1,1\right\}.\label{tau_d_p2}
\end{align}
\end{small}
\!\!For general $1\!\le\!r\!\le\!M$, the achievable communication latency pair is given by the lower convex envelope of the above points $\left\{\left(\tau^u_a(r),\tau^d_a(r)\right)\!:r\!\in\![M]\right\}$.
\end{theorem}
The achievable task offloading scheme for Theorem \ref{theorem3} is given in Section \ref{partialoffloadingscheme}.
\begin{theorem}\label{theorem4}
(Converse). The optimal communication latency pair $\left(\tau^{u^*}(r),\tau^{d^*}(r)\right)$ at any given computation load $r\in[1,M]$, for partial task offloading in the MEC network with $M$ ENs and $N$ users, is lower bounded by
\begin{small}
\begin{align}
&\tau^{u^*}(r)\ge\max\left\{\frac{Nr}{M},1\right\},\\
&\tau^{d^*}(r)\ge \frac{N}{\min\{M,N\}}.
\end{align}
\end{small}
\end{theorem}
\begin{figure*}
\vspace{-5mm}
\centering
\subfigure[$M=N=3$.]
{\label{innerouter2} 
 \includegraphics[width=2.7in, height=2.1in]{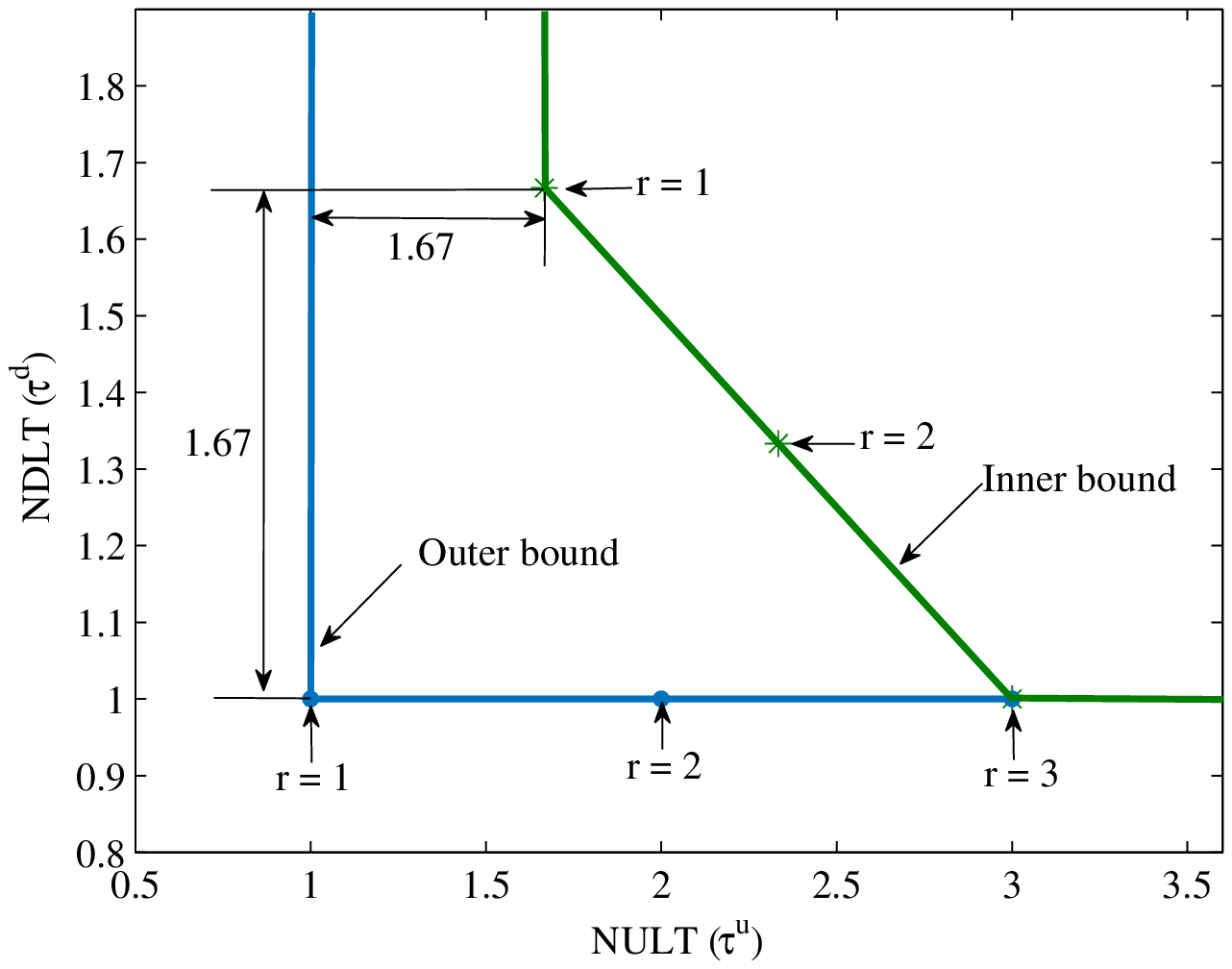}}
\hspace{12mm}
\subfigure[$M=N=10$.]
{\vspace{-3mm}\label{loadfunction11} 
 \includegraphics[width=2.7in, height=2.1in]{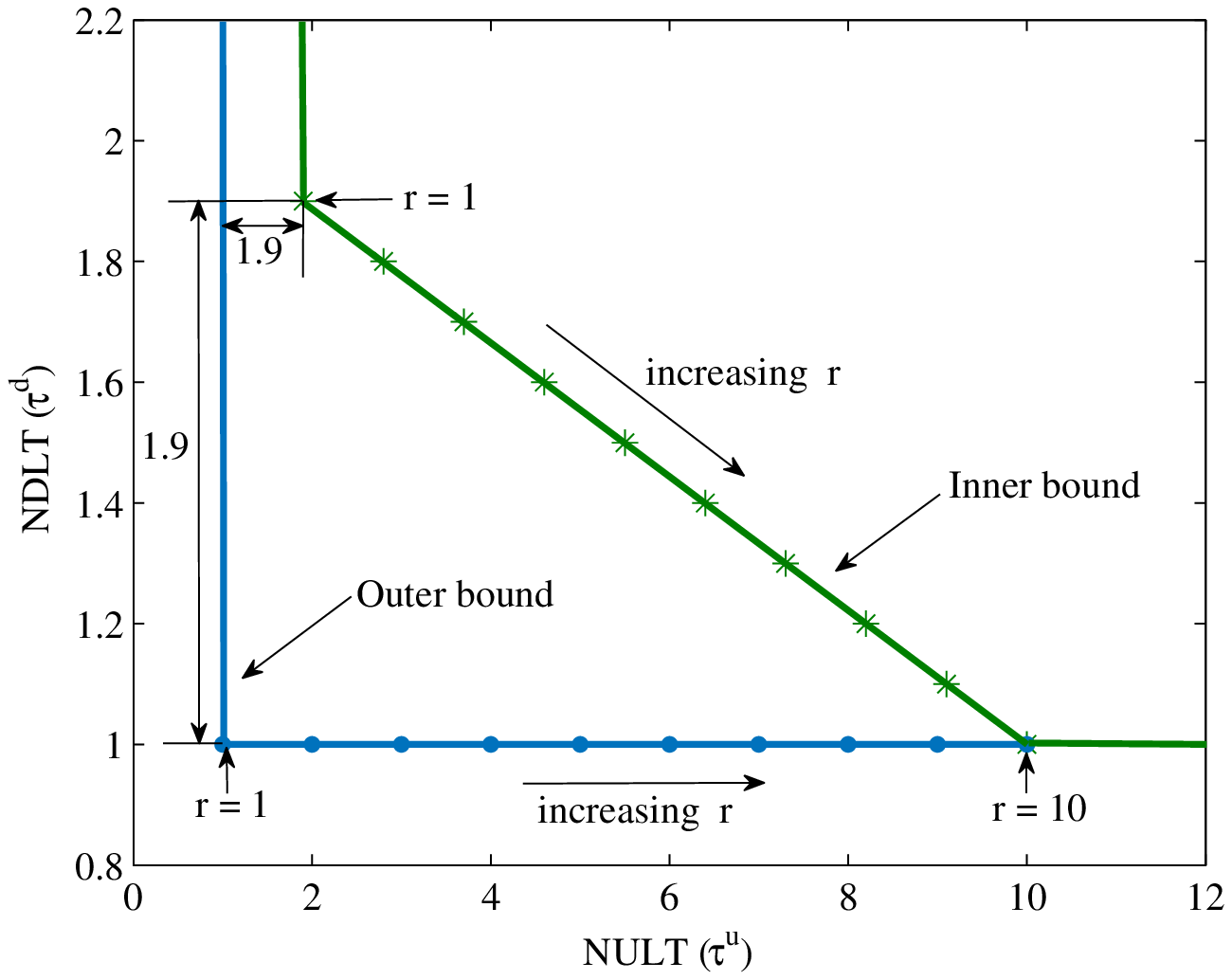}}
 \vspace{-2mm}
 \caption{The inner bound and outer bound of the optimal communication latency region for partial offloading. At a given computation load $r\in[1,M]$, both the gaps of NULT and NDLT are within $2$.}
\label{tradeoff_1_2} 
\vspace{-6mm}
\end{figure*}
According to Theorem \ref{theorem3} and Theorem \ref{theorem4}, we can obtain an inner bound $\mathscr{T}_{in}$ and an outer bound $\mathscr{T}_{out}$, respectively, of the optimal communication latency region by collecting the latency pairs $(\tau^u(r), \tau^d(r))$ at all the computation loads $r$'s. Fig. \ref{tradeoff_1_2} presents the bounds in the MEC networks with $M\!=\!N\!\in\!\{3,10\}$.
\begin{corollary}\label{corollary3}
(Optimality). At a computation load $r\in[1,M]$, both the achievable NULT in (\ref{tau_u_p2}) and NDLT in (\ref{tau_d_p2}) are within multiplicative gaps of $2$ to their lower bounds.
\end{corollary}
The proofs for Theorem \ref{theorem4} and Corollary \ref{corollary3} are given in Section \ref{partialconverse}.

Observing (\ref{tau_u_p2}) and (\ref{tau_d_p2}), we have
the monotonicity of the computation-communication function $\left(\tau^u_a(r),\tau^d_a(r)\right)$:
\begin{itemize}
\item The NULT $\tau^u_a(r)$ increases with the computation load $r$ strictly for $1\!\le\! r\!\le\! M$.
\item The NDLT $\tau^d_a(r)$ \textbf{decreases linearly} with the computation load $r$ for $1\!\le\! r\!\le\! \min\{M,N\}$, and keeps a constant $1$ for $\min\{\!M,N\}\!\le\! r\!\le\! M$.
\end{itemize}
Then, we have the following remark to characterize the inner bound of the optimal communication latency region, present the computation-communication tradeoff, and show the interaction between NULT and NDLT.
\begin{remark}
The inner bound $\mathscr{T}_{in}$ of the optimal communication latency region of partial task offloading is composed of two sections corresponding to two different intervals of the computation load $r$:
\begin{enumerate}
\item NULT-NDLT tradeoff section: $\tau^u(r)\!=\!\frac{N-1}{M}r+1$, $\tau^d(r)\!=\!\frac{N-r}{M}+1$, when $1\!\le \!r\!\le\! \min\{M,N\}$;
\item Constant-NDLT section: $\tau^u(r)\!=\!\frac{N-1}{M}r+1$, $\tau^d(r)\!=\!1$, when $\min\{M,N\}\!\le\! r\!\le \!M$.
\end{enumerate}
In particular, in the NULT-NDLT tradeoff section, the NDLT decreases linearly with the computation load, at the expense of increasing the NULT linearly.
\end{remark}
It is clearly seen from Fig. \ref{tradeoff_1_2} that in the $M\!=\!N$ case, a linear tradeoff between NDLT and NULT is established when the computation load increases within $[1,M]$.
\vspace{-3mm}
\subsection{Achievable task offloading scheme for Theorem \ref{theorem3}}\label{partialoffloadingscheme}
\subsubsection{Task partition and uploading}
Our task partition scheme is motivated by the transmitter caching schemes in\cite{7857805,FanXu} that utilize the duplicate caching of subfiles to enable transmitter cooperation such as zero-forcing and interference alignment. All tasks are treated equally in this work without taking user priority into account. We thus focus on the task partition of $W_j$ without loss of generality, for $j \!\in\! \mathcal{N}$. We partition $W_j$ into $\binom{M}{r}$ equal-sized subtasks, each denoted as $W_{j,\Phi}$ and to be computed at the ENs in subset $\Phi\!\subseteq\!\mathcal{M}$, with \emph{repetition order} $|\Phi|\!\!=\!r$. By Definition \ref{defenition1}, the computation load is calculated as $\frac{N\!M\binom{M-1}{\!r-1\!}L/\binom{\!M\!}{\!r\!}}{N\!L}\!=\!r$, which equals the \emph{repetition order} of subtasks. The output data size of a subtask is assumed to be proportional to its input data size.

According to the above task partition scheme, each user uploads the subtasks to the assigned ENs via uplink channels.
Specifically, each EN $i$ wants subtasks $\{W_{j, \Phi}\!:\!\forall j\!\in\! \mathcal{N}, \forall \Phi\!\supseteq\!\{i\}, |\Phi|\!=\!r\}$. There are a total of $N\binom{M}{r}$ subtasks in this transmission at the computation load $r$, and each EN desires $N\binom{M-1}{r-1}$ subtasks of them. 
Clearly, the channel formed by uploading the subtasks at the computation load $r$ is the \emph{X-multicast channel with multicast group size} $r$, defined in \cite{FanXu,DOfNiesen}. The optimal per-receiver DoF of this channel is given below via interference alignment.
\begin{corollary} \label{corolary 1}
(\cite[Lemma 1]{FanXu}, \cite[Theorem 2]{DOfNiesen}) The optimal per-receiver DoF of the X-multicast channel with $N$ transmitters, $M$ receivers, and multicast group size $r$ is given by
\begin{small}
\begin{equation}
DoF^{u}_{r}=\frac{Nr}{Nr+M-r}, ~~~r\in[M].  \label{dofu}
\end{equation}
\end{small}
\end{corollary}
The traffic load for each EN to receive its assigned subtasks is $N\binom{M-1}{r-1}\frac{L}{\binom{M}{r}}\!=\!\frac{Nr}{M}L$ bits. Similar to (\ref{T^{u'}(m)}) in binary offloading, the NULT for each EN at computation load $r$ is given by
\begin{small}
\begin{equation}  \label{tau_u_p}
\tau^u_a(r)=\frac{\frac{Nr}{M}}{DoF^{u}_{r}}=\frac{N-1}{M}r+1.
\end{equation}
\end{small}
\vspace{-5mm}
\subsubsection{Results downloading}
After computing all the offloaded subtasks, ENs begin to transmit the computed results back to users via downlink channels. In specific, each user $j$ wants the results of subtasks $\{W_{j, \Phi}\!: \forall\Phi\!\subseteq\!\mathcal{M},|\Phi|\!=\!r\}$. There are a total of $N \binom{M}{r}$ subtasks in this transmission at the computation load $r$, of which $\binom{M}{r}$ subtasks are desired by each user $j$. The duplication of computed results on multiple ENs can enable the transmitter cooperation to be exploited in results downloading. The downlink channel formed by downloading the computed results at the computation load $r$
is the \emph{cooperative X channel with transmitter cooperation group size} $r$ defined in \cite{FanXu,TxCache}. An achievable per-receiver DoF of this channel is given below.
\begin{corollary}\label{corollary2}
(\cite[Theorem 1]{TxCache}) An achievable per-receiver DoF of the cooperative X channel with $M$ transmitters, $N$ receivers, and transmitter cooperation group size $r$ is given by
\begin{small}
\begin{equation}\label{dofd}
DoF^{d}_{r}=\min\left\{\frac{M}{M+N-r},1\right\},~~r\in[M],
\end{equation}
\end{small}
and it is within a multiplicative gap of $2$ to the optimal DoF.
\end{corollary}
The traffic load for each user to download its task output data is $\widetilde{L}$ bits. Similar to (\ref{T^{d'}_m}) in binary offloading, the NDLT for each user at computation load $r$ is given by
\begin{small}
\begin{equation} \label{tau_d_p}
\tau^d_a(r)= \frac{1}{DoF^{d}_{r}}=\max\bigg\{\frac{N-r}{M}+1,1\bigg\},~~r\in[M].
\end{equation}
\end{small}
\!\!By (\ref{tau_u_p}) and (\ref{tau_d_p}), we thus obtain an achievable communication latency pair $\left(\tau^u_a(r),\tau^d_a(r)\right)$ at an integer computation load $r\!\in\![M]$ for partial offloading.
\subsubsection{Non-integer computation load}
Based on the above scheme, given an integer computation load $r\!\in\![M]$, the achievable NULT $\tau^u_a(r)$ and NDLT $\tau^d_a(r)$ are given by (\ref{tau_u_p}) and (\ref{tau_d_p}), respectively. If $r$ is a non-integer value, it can be rewritten as as a convex combination
of $\lceil r\rceil$ and $\lfloor r\rfloor$, i.e., $r\!=\!\lambda\lceil r\rceil\!+\! (1\!-\!\lambda)\lfloor r\rfloor$ for some $\lambda\!\in\![0,1]$. Then, we can partition the input data of each task into two parts with ratio $\lambda$ and $1\!-\!\lambda$, respectively. Then, we apply the achievable schemes with computation load $\lceil r\rceil$ and $\lfloor r\rfloor$ to offload these two parts, respectively, via a time-sharing way. The corresponding communication latency pair is given by $\tau^{u}_a(r) \!=\! \lambda \tau^{u}_a(\lceil r\rceil) \!+\! (1\!-\!\lambda)\tau^{u}_a(\lfloor r\rfloor)$, and $\tau^{d}_a(r) \!=\! \lambda \tau^{d}_a(\lceil r\rceil) \!+ \! (1\!-\!\lambda)\tau^{d}_a(\lfloor r\rfloor)$, i.e., $\left(\tau^{u}_a(r),\tau^{d}_a(r)\right)\! = \!\lambda \left(\tau^{u}_a(\lceil r\rceil),\tau^{d}_a(\lceil r\rceil)\right) \!+ (1\!-\!\lambda)\!\left(\tau^{u}_a(\lfloor r\rfloor),\tau^{d}_a(\lfloor r\rfloor)\right)$. Actually, for any two integer-valued computation loads $r_1$ and $r_2$, the points on the line
segment between $(\tau^{u}_a(r_1),\tau^{d}_a(r_1))$ and $(\tau^{u}_a(r_2),\tau^{d}_a(r_2))$ can be achieved via data partition and time sharing. So the lower convex envelope of achievable points $\left\{\left(\tau^u_a(r),\tau^d_a(r)\right)\!: r\!\in\![M]\right\}$ is also achievable.
\vspace{-5mm}
\subsection{Proof of Converse for Theorem \ref{theorem4} and Corollary \ref{corollary3}} \label{partialconverse}
\subsubsection{Lower bound and gap of NULT}
This proof is similar to the lower bound proof of the NULT for binary offloading. First, we derive an lower bound on the NULT of any given feasible subtask assignment policy with computation load $r\!\in\![1,M]$. Then, we construct an optimization problem to obtain the lower bound of the NULT for all feasible subtask assignment policies.

Given a computation load $r\!\in\![1,M]$, consider an arbitrary subtask assignment $\left\{W_{j,\Phi}\right\}$, the subtasks partitioned from task $W_j$ and assigned to EN $i$ are denoted as $\mathcal{W}_{j,i}\!\triangleq\!\left\{W_{j, \Phi}\!: \forall \Phi\!\supseteq\!\{i\}\right\}$, $\forall i\!\in\!\mathcal{M}$ and $\forall j\!\in\!\mathcal{N}$, and the size of $\mathcal{W}_{j,i}$ is assumed to be $\gamma_{j,i}L$, and $\gamma_{j,i}$ satisfies
\begin{small}
\begin{align}
&\sum\nolimits_{i\in\mathcal{M}}\gamma_{j,i} = r ,~~\forall j \in\mathcal{N} \label{cons111}\\
&~0\le \gamma_{j,i} \le 1,~~\forall j \in\mathcal{N},~\forall i\in\mathcal{M}. \label{cons222}
\end{align}
\end{small}
\!\!Let $\mathbf{y}_i$ denote the signal received at each EN $i$, and $\mathbf{x}_j$ the signal transmitted by each user $j$, over the block length $T^u$.
Hence, for any EN $i \in\mathcal{M}$, we have the following chain of inequalities,
\begin{small}
\begin{align}
\sum\nolimits_{j \in\mathcal{N}}\gamma_{j,i} L &= H(\mathcal{W}_{1,i},\cdots,\mathcal{W}_{N,i})\\
&=I(\mathcal{W}_{1,i},\cdots,\mathcal{W}_{N,i};\mathbf{y}_i) + H(\mathcal{W}_{1,i},\cdots,\mathcal{W}_{N,i}|\mathbf{y}_i)\\
&\le I(\mathcal{W}_{1,i},\cdots,\mathcal{W}_{N,i};\mathbf{y}_i) + \sum\nolimits_{j \in\mathcal{N}}H(\mathcal{W}_{j,i}|\mathbf{y}_i)\\
&\le I(\mathbf{x}_1,\mathbf{x}_2,\cdots,\mathbf{x}_N;\mathbf{y}_i) + NT^u\epsilon \\
&\le T^u \log P_u  + NT^u\epsilon.
\end{align}
\end{small}
\!\!By dividing on $\frac{L}{\log P_u}$, and taking $P_u\!\to\!\infty$ and $\epsilon\!\to\!0$, we have
$\tau^u \ge \sum_{j\in\mathcal{N}}\gamma_{j,i}$.

On the other hand, for any user $j\!\in\!\mathcal{N}$ and its task input data $W_j\!=\!\cup_{i\in\mathcal{M}}\mathcal{W}_{j,i}$, we also have
\begin{small}
\begin{align}
L \!=\!H(W_j)&\!=\! H(\mathcal{W}_{j,1},\cdots,\mathcal{W}_{j,M}) \\
\!&=\! I(\mathcal{W}_{j,1},\cdots,\mathcal{W}_{j,M};\mathbf{y}_1,\cdots,\mathbf{y}_M) \!+\! H(\mathcal{W}_{j,1},\cdots,\mathcal{W}_{j,M}|\mathbf{y}_1,\cdots,\mathbf{y}_M) \\
\!&\le\! I(\mathbf{x}_j\!;\!\mathbf{y}_1,\cdots,\mathbf{y}_M) \!+\! \sum\nolimits_{i\in\mathcal{M}}H(\mathcal{W}_{j,i}|\mathbf{y}_i)\\
\!&\le\! T^u \log P_u \!+\! MT^u\epsilon.
\end{align}
\end{small}
\!\!\!By dividing on $\frac{L}{\log P_u}$, and taking $P_u\!\to\!\infty$ and $\epsilon\!\to\!0$, we have $\tau^u\!\ge\! 1$. Combining these two final inequalities of $\tau^u$, for any given feasible subtask assignment $\boldsymbol{\gamma}\triangleq[\gamma_{j,i}]_{i\in\mathcal{M},\forall j\in\mathcal{N}}$, the NULT satisfies $\tau^u \ge \max\left\{\sum_{j\in\mathcal{N}}\gamma_{j,i},1\right\}$, $\forall i\in\mathcal{M}$, i.e., the minimum NULT of the subtask assignment policy $\boldsymbol{\gamma}$ is lower bounded by
\begin{small}
\begin{equation}
\!\!\!\!\tau^{u^*}\!(r,\boldsymbol{\gamma}) \!\ge\! \max\limits_{ i\in\mathcal{M}} \max\!\bigg\{\!\sum_{j\in\mathcal{N}}\gamma_{j,i},1\!\bigg\}\!=\!\max\bigg\{\!\max\limits_{ i\in\mathcal{M}}\sum_{j\in\mathcal{N}}\gamma_{j,i},1\bigg\}.\!
\end{equation}
\end{small}
\!\!Hence, the minimum NULT of all subtask assignment policies is given by $\tau^{u^*}(r)\!=\!\min\limits_{\boldsymbol{\gamma}}\tau^{u^*}(r,\boldsymbol{\gamma})$, and can be lower bounded by the optimal solution of the following linear programming problem,
\begin{small}
\begin{align}
\mathrm{P2}:\quad&\min\limits_{\boldsymbol{\gamma}}~\max\!\bigg\{\max\limits_{ i\in\mathcal{M}}\sum_{j \in\mathcal{N}}\gamma_{j,i},1\bigg\} \nonumber \\
&~\mathnormal{s.t.}\quad(\ref{cons111}), (\ref{cons222})\nonumber
\end{align}
\end{small}
\!\!\!By defining a new variable $\mu_{i}=\sum_{j\in\mathcal{N}}\gamma_{j,i}$, Problem $\mathrm{P2}$ can be transformed into $\mathrm{P3}$ below,
\begin{small}
\begin{align}
\mathrm{P3}:\quad&\min\limits_{\boldsymbol{\mu}}~\max\!\Big\{\max\limits_{ i\in\mathcal{M}}\mu_{i},1\Big\} \nonumber\\
&~\mathnormal{s.t.}\quad\sum_{i \in\mathcal{M}}\mu_{i} = Nr ,  \\
                 &~~~~~~\quad0\le \mu_{i} \le N,~~\forall i \in\mathcal{M}.
\end{align}
\end{small}
\!\!For any given feasible solution $\{\mu_{i}\}$ to $\mathrm{P3}$, we can also construct a feasible solution to $\mathrm{P2}$ by let $\gamma_{j,i}\!=\!\frac{\mu_{i}}{N}$, $\forall j \!\in\!\mathcal{N}$ and $\forall i\!\in\!\mathcal{M}$. Thus, Problem $\mathrm{P2}$ and $\mathrm{P3}$ are equivalent. It can be easily proved that the optimal solution to $\mathrm{P3}$ is given by $\mu^*_i\!=\!\frac{Nr}{M}$, and consequently, the optimal solution to $\mathrm{P2}$ is $\gamma^*_{j,i}\!=\!\frac{r}{M}$. Therefore, the minimum NULT $\tau^{u^*}(r)$ is lower bounded by
$\tau^{u^*}(r)\!\ge\! \max\left\{\frac{Nr}{M},1\right\}. \label{partial_lowerboundtau}$
Comparing it with the achievable NULT in (\ref{tau_u_p2}) of Theorem \ref{theorem3}, it can be easily proved that the multiplicative gap between them is within 2, i.e.,
$\frac{\tau^u_a(r)}{\tau^{u^*}(r)}\!\le\!\frac{\frac{N-1}{M}r+1}{\max\{ Nr/M,1\}}\le 2$.
Thus, we complete the proof of the lower bound and gap of the NULT for partial offloading.
\subsubsection{Lower bound and gap of NDLT}
Since the downlink channel capacity in this problem cannot exceed the capacity of the $N$ user MISO broadcast channel with a single $M$-antenna transmitter, so the NDLT can be lower bounded by
$\tau^{d^*}\!\ge\!\frac{N}{\min\{M,N\}}. \label{partial_lowerboundtaud} $
The rigorous proof is the same as the lower bound proof of NDLT for binary offloading in Section \ref{binaryconversedown}. The multiplicative gap between the achievable NDLT in (\ref{tau_d_p2}) of Theorem \ref{theorem3} and the above lower bound is within $2$, i.e.,
$\frac{\tau^{d}_a(r)}{{\tau^{d}}^{*}(r)} \le \frac{\max\left\{\frac{N-r}{M}+1,1\right\}}{N/\min\{M,N\}}\le 2$.
\vspace{-3mm}
\section{Conclusion} \label{conclusion}
\vspace{-1mm}
This paper studied a fundamental tradeoff between computation load $r$ and communication latency $\left(\tau^u(r),\tau^d(r)\right)$ defined as the pair of normalized uploading time (NULT) and normalized downloading time (NDLT), in the MEC network with $M$ ENs and $N$ users. We exploited the idea of computation replication in task offloading schemes to speed up the computed result downloading via transmission cooperation, in binary and partial offloading cases. We developed an order-optimal achievable communication latency pair at a given computation load, and both the NULT and NDLT are within multiplicative gaps of $2$ to their lower bounds. Particularly, the NULT in binary offloading is optimal. We showed that the NDLT can be traded by the computation load $r$ in the specific interval. It is an \emph{\textbf{inversely proportional function}} for $\frac{MN}{M\!+\!N}\!\le\! r \!\le\! M$ in binary offloading, and a \emph{\textbf{linear decreasing function}} for $1\!\le\! r\!\le \!\min\{M,N\}$ in partial offloading, both of which decrease at the expense of increasing the NULT linearly. Hence, computation replication is very beneficial in reducing the communication latency for offloading tasks whose output data size is larger than the input data size. Our results revealed a fundamental relationship between computation load and communication latency in MEC systems. Numerical examples also verify the significant gain brought by computation replication on reducing the end-to-end task execution time in the high SNR regime.

This paper is devoted to providing first-order insights into the communication-computation tradeoff in MEC systems from the information-theoretic perspective. The theoretical analysis relies on several ideal assumptions on both the computing model and the communication model. Future works can consider more practical scenarios where the ENs may have different computing capabilities, some offloaded tasks may be inter-dependent, the users and ENs are partially connected, and so so.
\vspace{-4mm}
\appendix
\vspace{-3mm}
\subsection{Proof of Lemma \ref{lemma4}}\label{lemma4proof}
\subsubsection{Achievable scheme}\label{achievable1}We use the partial interference alignment scheme with a $u_s \!=\!n s^\Gamma \!+(s+1)^{\Gamma}$ symbol extension over the original channel, where $s\!\in\!\mathbb{N}$ and $\Gamma \!=\! M(N\!-n)$. Specifically, each transmitter $j$ encodes the task input message $W_j$ into $s^\Gamma$ independent streams $x^{l}_{j}$, $l\!\in\![s^\Gamma]$, each beamformed along a $u_s\!\times\!1$ column vector $\mathbf{v}^{l}_{j}$. So the symbol $\bar{\mathbf{X}}_{j}$ transmitted at transmitter $j$ can be expressed as
\begin{small}
\begin{equation}
\bar{\mathbf{X}}_{j} = \sum_{l=1}^{s^\Gamma}x^{l}_{j} \mathbf{v}^{l}_{j}=\bar{\mathbf{V}}_{j}\mathbf{X}_{j},
\end{equation}
\end{small}
\!\!where $\mathbf{X}_{j}\!\triangleq\!(x^{l}_{j})^{s^\Gamma}_{l=1}$ is a $s^\Gamma\!\times\!1$ column vector, and $\bar{\mathbf{V}}_{j}\!=\![\mathbf{v}^{l}_{j} ]^{s^\Gamma}_{l=1}$ is a
$u_s\!\times\!s^\Gamma$ matrix. Then, the received signal at EN $i$ can be written as
\begin{small}
\begin{equation}
\bar{\mathbf{Y}}_{i}=\sum_{j=1}^{N}\bar{\mathbf{H}}_{ij}\bar{\mathbf{V}}_{j}\mathbf{X}_{j}+\bar{\mathbf{Z}}_{i},
\end{equation}
\end{small}
\!\!where $\bar{\mathbf{Y}}_{i}$ and $\bar{\mathbf{Z}}_{i}$ represent the $u_s$ symbol extension of the received signal $Y_{i}$ and noise $Z_{i}$, respectively. $\bar{\mathbf{H}}_{ij}$ is a $u_s\!\times\!u_s$ diagonal matrix
representing the $u_s$ symbol extension of the channel, whose $l$-th diagonal element is $h_{ij}(l)$.

Next, we design the beamforming vectors such that each receiver $i$ can decode the $n$ desired signals $\{\mathbf{X}_{k+\!1}\!:\! k\!\in\![(i\!-\!1)n\!:\!(in\!-\!1)]\!\!\!\pmod{\!N}\}$ by zero-forcing the interferences. To align the $N\!-\!n$ interference signals at each receiver together in the space with dimension $(s\!+\!1)^{\Gamma}$,  beamforming vectors need to satisfy the following conditions:
\begin{equation}
\text{span}(\bar{\mathbf{H}}_{ij}\!\bar{\mathbf{V}}_{j})\!\prec\!\text{span}(\mathbf{V}),\forall i\!\in\!\mathcal{M},\label{alignment}
\end{equation}
for $\forall j\!\in\!\mathcal{N},(i,j)\!\notin\! \left\{\left(i, k\!+\!\!1\right)\!:\!k\!\in\![(i\!-\!\!1)n\!:\!(in\!-\!\!1)]\!\!\!\pmod{\!\!N}\right\}$, where span($\mathbf{P}$) denotes the space spanned by the column vectors of matrix $\mathbf{P}$, and $\mathbf{V}$ is a $u_s\!\times \!(s+\!1)^\Gamma$ matrix. Now we need to design the column vectors
of $ \bar{\mathbf{V}}^{[j]}$ and $\mathbf{V}$ to satisfy (\ref{alignment}). Let $\mathbf{w}$ be a $u_s \!\times\! 1$ column vector $\mathbf{w} \!=\! (1,1,\cdots,1)^{\mathrm{ T }}$. The sets of column vectors of $ \bar{\mathbf{V}}_{j}$ and $\mathbf{V}$, denoted as $ \bar{\mathcal{V}}_{j}$ and $\mathcal{V}$, respectively, are given as below
\begin{small}
\begin{align}
&\bar{\mathcal{V}}_{j}\!=\! \underbrace{\!\left\{\!\!\!\left(\prod_{t\in\!\mathcal{M},q\in\!\mathcal{N}\!,(t,q)\notin\left\{\!\left(t, k+\!1\right):k\in[(\!t\!-\!1\!)n:(\!tn\!-\!1\!)]\!\!\!\!\!\!\pmod{\!\!N}\!\right\}}\!\!\!\!\!\!\!\!\!\!\!\!\!\!\left(\bar{\mathbf{H}}_{tq}\!\right)^{\!\alpha_{tq}} \!\!\!\right)\!\!\mathbf{w}\!:\!\alpha_{tq}\!\!\in\![0\!:\!s\!-\!\!1]\!\!\right\}}_{\text{a total of } s^{\Gamma} \text{ columns}}\label{v1}
\end{align}
for $\forall j\!\in\!\mathcal{N}$, and
\begin{align}
&\!\mathcal{V}\!=\! \underbrace{\!\!\left\{\!\!\!\left(\prod_{t\in\!\mathcal{M},q\in\mathcal{N}\!,(t,q)\notin\left\{\!\left(\!t, k+\!1\!\right):k\in[(\!t-\!1\!)n:(\!tn-\!1\!)]\!\!\!\!\!\!\pmod{\!\!N}\!\right\}}\!\!\!\!\!\!\!\!\!\!\!\left(\bar{\mathbf{H}}_{tq}\right)^{\alpha_{tq}} \!\!\right)\!\!\mathbf{w}\!:\!\alpha_{tq}\!\!\in\!\![0\!:\!s]\!\!\right\}}_{\text{a total of } (s+1)^{\Gamma} \text{ columns}}\!\!.\label{v2}
\end{align}
\end{small}
\!\!\!We then show that the desired signal streams received at each receiver are linearly independent of each other and interference signal streams such that the desired streams can be decoded by zero-forcing
interferences. At any receiver $i$, the desired signal streams are beamformed along the $n s^\Gamma$ vectors of $\left[\bar{\mathbf{H}}_{i,i_1}\!\bar{\mathbf{V}}_{i_1}~~\bar{\mathbf{H}}_{i,i_2}\!\bar{\mathbf{V}}_{i_2}~\cdots~\bar{\mathbf{H}}_{i,i_n}\!\bar{\mathbf{V}}_{i_n}\right]$, where $i_m \!=\! (i\!-\!1)n\!+\!m\!\!\pmod{\!N}$, $m\!\in\![n]$.
By condition (\ref{alignment}), the interference streams at any receiver $i$ from transmitter $j$ are aligned at the column vector space of $\mathbf{V}$ for $j\in[N]/\left\{k\!+\!1\!:\!k\!\in\![(i\!-\!1)n\!:\!in\!-\!1]\!\!\pmod{\!N}\right\}$. To decode the desired $n s^\Gamma$ streams successfully, it suffices to show that the $u_s\!\times\!u_s$ matrix
\begin{small}
\begin{equation}
\mathbf{A}_i=\left[\bar{\mathbf{H}}_{i,i_1}\!\bar{\mathbf{V}}_{i_1}~~\bar{\mathbf{H}}_{i,i_2}\!\bar{\mathbf{V}}_{i_2}~\cdots~\bar{\mathbf{H}}_{i,i_n}\!\bar{\mathbf{V}}_{i_n} ~~ \mathbf{V}\right]
\label{signal_space}
\end{equation}
\end{small}
\!\!has a full rank of $u_s$ almost surely for $i\in\mathcal{M}$. By the beamforming vectors in (\ref{v1}) and (\ref{v2}), we can observe that the $u_s$ elements in the $l$-th row of $\mathbf{A}_i$ have the following forms
\begin{small}
\begin{equation}
\underbrace{\left\{\!h_{i,i_{m}}\!(l)\!\!\!\!\!\!\!\!\!\!\!\!\!\!\!\!\!\prod_{t\in\mathcal{M},q\in\mathcal{N}\!,(\!t,q\!)\notin \left\{\!\left(\!t, k+\!1\!\right):k\in[(\!t-\!1\!)n:(\!tn-\!1\!)]\!\!\!\!\!\!\pmod{\!\!N}\!\right\}}\!\!\!\!\!\!\!\!\!\!\!\!\!\!\!\!\!\!\!\!\!\!\!\!\!\!\!\!\!\!\!\!\left(
h_{tq}(l)\!\right)^{\alpha_{tq}}\!\!:\!\alpha_{tq}\!\!\in\!\![0\!:\!\!s\!-\!\!1],\!m\!\!\in\!\![n]\!\!\right\}\!}_{\text{a total of } ns^{\Gamma} \text{ elements}}\bigcup\label{binoim1} \nonumber
\end{equation}
\begin{equation}
\!\!\underbrace{\!\left\{\!\prod_{t\in\mathcal{M},q\in\mathcal{N}\!,(t,q)\notin \left\{\!\left(\!t, k+\!1\!\right):k\in[(\!t-\!1\!)n:(\!tn-\!1\!)]\!\!\!\!\!\!\pmod{\!\!N}\!\right\}}\!\!\!\!\!\!\!\!\left( h_{tq}(l)\!\right)^{\beta_{tq}}\!\!:\!\beta_{tq}\!\!\in\![0\!:\!s]\!\right\}}_{\text{a total of } (s+1)^{\Gamma} \text{ elements}},    \label{binoim2}
\end{equation}
\end{small}
\!\!\!where $\{h_{ij}(l)\}$ are drawn independently from a continuous probability distribution. We observe from (\ref{binoim2}) that all the elements of $\mathbf{A_i}$ meet the two conditions of \cite[Lemma 1]{Xdof}. Hence, the matrix $\mathbf{A_i}$, $\forall i\in\mathcal{M}$, is a full-rank matrix. Taking $s$ to infinity, the DoF for each receiver achieved by above scheme is given by
${\lim_{s \to +\infty} \frac{ns^\Gamma}{n s^\Gamma + (s+1)^{\Gamma}}}\!=\!\frac{n}{n+1}\!=\!\frac{Nr}{Nr+M}$.

Further, consider the basic scheme that $N$ transmitters deliver their messages to the assigned receivers in the time division manner, which achieves a DoF of $\frac{n}{N}\!=\!\frac{r}{M}$ for each receiver. Therefore, the per-receiver DoF of the considered interference-multicast channel with multicast group size $r$ is given by $DoF^{u}_r\!=\!\max\!\big\{\!\frac{Nr}{Nr+M},\frac{r}{M}\!\big\} $.
\subsubsection{Converse Proof of Lemma \ref{lemma4}}\label{converseup}
In Section \ref{binaryconverseup}, we have proved that the NULT of the task assignment and uploading scheme in Section \ref{binary_schemeuploading} is information-theoretically optimal, so the per-receiver DoF of the considered uplink channel in (\ref{dof_u}) of Lemma \ref{lemma4} must also be optimal.

%
\vspace{-4mm}
\subsection{Proof of Lemma \ref{lemma5}}\label{lemma5proof}
\subsubsection{Achievability} We show the achievable schemes in two cases, $r=1$ and $r\ge 2$, respectively.\\
$i)$ $r\!=\!1$:
By (\ref{ntask}), the task output messages at each transmitter $i$ can be represented as  $\big\{\widetilde{W}_{j}\!:\!j\!\in\![(i\!-\!1)n\!+\!1\!:\!in]\!\big\}$.
Let $\Gamma \!=\! M(N\!-\!1)$ and consider a $u_s \!= \!s^\Gamma \!+\! n(s\!+\!1)^{\Gamma}$ symbol extension.
Each message $\widetilde{W}_{j}$ are encoded into $s^\Gamma$ independent streams $x^{l}_{j}$, $l\!\in\![s^\Gamma]$, each beamformed along a $u_s \!\times\! 1$ column vector $\mathbf{v}^{l}_{j}$, $\forall j\!\in\!\mathcal{N}$. Then, the signal transmitted at transmitter $i$ can be expressed as
\begin{small}\begin{equation}
\bar{\mathbf{X}}_{i} = \!\sum_{j=(i-1)n+1}^{in}\sum_{l=1}^{s^\Gamma}x^{l}_{j} \mathbf{v}^{l}_{j}=\!\sum_{j=(i-1)n+1}^{in}\!\!\!\bar{\mathbf{V}}_{j}\mathbf{X}_{j},
\end{equation}\end{small}
\!\!where $\mathbf{X}_{j}=(x^{l}_{j})^{s^\Gamma}_{l=1}$ is a $s^\Gamma\!\times\!1$ column vector and $\bar{\mathbf{V}}_{j}=[\mathbf{v}^{l}_{j}]^{s^\Gamma}_{l=1}$ is a $u_s\!\times\!s^\Gamma$ matrix. The signal received at receiver $j$ can be expressed as
\begin{small}\begin{equation}
\bar{\mathbf{Y}}_{j}=\sum_{i=1}^{M}\bar{\mathbf{G}}_{ji}\!\sum_{k=(i-1)n+1}^{in}\!\!\!\bar{\mathbf{V}}_{k}\mathbf{X}_{k}+\bar{\mathbf{Z}}_{j},
\end{equation}\end{small}
\!\!where $\bar{\mathbf{G}}_{ji}$ is a $u_s \times u_s$ diagonal matrix representing the $u_s$ symbol extension of the channel, $\bar{\mathbf{Y}}_{j}$ and $\bar{\mathbf{Z}}_{j}$ represent the $u_s$ symbol extension of the received signal $Y_{j}$ and noise $Z_{j}$, respectively.

Next, we align the interferences at each receiver $j$ such that the total dimension of the spaces spanned by the interference vectors is $n (s+\!1)^{\Gamma}$. Then, the desired $ s^\Gamma$ streams corresponding to the desired signal $\mathbf{X}_j$ can be decoded by zero-forcing the interferences from
an $u_s \!=\! s^\Gamma \!+n(s+\!1)^{\Gamma}$-dimensional received signal vector. We ensure this by designing the beamforming vectors $\left\{\!\bar{\mathbf{V}}_{j}\!\right\}$ as follows, where the message $\widetilde{W}_j$ desired by receiver $j$ is at transmitter $\lceil \frac{j}{n}\rceil\!\in\!\mathcal{M}$ by (\ref{ntask}),
\begin{small}\begin{equation}\label{aligneq2}
\left.
\begin{aligned}
\text{span}(\bar{\mathbf{G}}_{ji}\bar{\mathbf{V}}_{\!(i-1)n+1})&\subset \text{span}(\mathbf{U}_{1})\\
\text{span}(\bar{\mathbf{G}}_{ji}\bar{\mathbf{V}}_{\!(i-1)n+2})&\subset \text{span}(\mathbf{U}_{2})\\
&\vdots\\
\text{span}(\bar{\mathbf{G}}_{ji}\bar{\mathbf{V}}_{\!(i-1)n+k})&\subset \text{span}(\mathbf{U}_{k})\\
&\vdots\\
\text{span}(\bar{\mathbf{G}}_{ji}\bar{\mathbf{V}}_{\!in})&\subset \text{span}(\mathbf{U}_{n})
\end{aligned}
\!\right\}
\begin{aligned}
\!\forall j\!\in\!\mathcal{N},\forall i\!\in\!\mathcal{M}, 
 j\!\!\ne\!\!(i\!-\!1\!)n\!+\!k~ \text{for}~ \forall k\!\in\! [n],
\end{aligned}
\end{equation}\end{small}
\!\!\!where $\mathbf{U}_{k}$ is a $u_s\!\times\!(s\!+\!1)^\Gamma$ matrix, $\forall k\!\in\![n]$. Next, we design $\left\{\mathbf{V}_{\!j}\right\}$ and $\left\{\mathbf{U}_{k}\right\}$ to satisfy above conditions. First, we generate $n$ $u_s\!\times\!1$ column vectors $\mathbf{w}_{k}\!=\!(w^{l}_{k})^{u_s}_{l=1},~k\!\in\![n]$. All elements of these $n$ vectors are chosen i.i.d from some continuous distribution whose
support lies between a finite minimum value and a finite maximum value. Then, the sets of column vectors of $ \bar{\mathbf{V}}_{\!j}$ and $\mathbf{U}_{k}$ are denoted as $\bar{\mathcal{V}}_{j}$ and $\mathcal{U}_{k}$, respectively, and are given as follows,
\begin{align}
\bar{\mathcal{V}}_{(i-\!1)n +k}&\!=  \! \left\{\!\!\left(\prod_{q\in\mathcal{N}\!,t\in\mathcal{M},(q,t)\ne(\!(t-\!1\!)n +k,t)}\!\!\!\!\!\!\!\!\left(\bar{\mathbf{G}}_{qt}\right)^{\!\alpha_{qt}}\!\!\right) \!\!\mathbf{w}_{k}\!: \!\alpha_{qt}\!\in\![0\!:\!s\!-\!\!1]\!\right\}\\
\!\!\!\!\!\!\!\!\!\!\!\!\text{for}~\forall i\!\in\!\mathcal{M},\forall k\!\in\![n],~\text{and}~~~~~~~~\nonumber\\
\mathcal{U}_{k}&\!= \!\left\{\!\!\left(\prod_{t\in\mathcal{M},q\in\mathcal{N},(q,t)\ne(\!(t-\!1\!)n +k,t)}\!\!\!\!\!\!\left(\bar{\mathbf{G}}_{qt}\right)^{\!\alpha_{qt}}\!\!\!\right)\!\!\mathbf{w}_{k}\!: \alpha_{qt}\!\in\![0\!:\!s]\!\right\}
\end{align}
for $\forall k\!\in\![n].$
In the following, we show that the desired signal streams are linearly
independent with the interference signal streams, and hence can be decoded by zero-forcing the
interference. Consider the signal vectors received at any receiver $j\!=\!(i\!-\!1)n\!+\!k$, $i\!\in\!\mathcal{M}$, $k\!\in\![n]$.
By (\ref{aligneq2}), the desired signal streams are beamformed along the $ s^\Gamma$ vectors of $\bar{\mathbf{G}}_{(i-\!1\!)n+k,i}\bar{\mathbf{V}}_{\!(i-\!1\!)n+k}$, while the interference vectors
are aligned at the column vector spaces of $\mathbf{U}_{k'},\forall k'\!\in\![n]$. To decode the desired streams successfully, it suffices to show that the $u_s \!\times\! u_s$ matrix
\begin{equation}
\mathbf{\Lambda}_{j}=\mathbf{\Lambda}_{(i-\!1\!)n+k}=\left[\bar{\mathbf{G}}_{(i-\!1\!)n+k,i}\!\bar{\mathbf{V}}_{\!(i-1)n+k}~~\mathbf{U}_{1}~~\mathbf{U}_{2}~\cdots~\mathbf{U}_{n}~\right] \label{signal_space}\end{equation}
is a full-rank matrix almost surely for $j\in\mathcal{N}$ or $\forall i\!\in\!\mathcal{M}$ and $\forall k\!\in\![n]$. It is seen that the $l$-th row elements of $\mathbf{\Lambda_{j}}$ have the following forms,
\begin{small}
\begin{align}
\!&\left\{h_{(\!i-\!1\!)n+k,i}(l)\!\!\!\!\!\!\!\prod_{q\in\mathcal{N}\!,t\in\mathcal{M},(\!q,t\!)\ne(\!(t-\!1\!)n +\!k,t)}\!\!\!\!\!\!\!\!\!\!\!\!\left( g_{qt}(l)\!\right)^{\!\alpha_{qt}}\!\!w_{k}(l)\!:\!\alpha_{qt}\!\!\in\!\![0\!:\!s\!-\!\!1]\!\right\}\!\bigcup\nonumber\\
\!&\left\{\prod_{q\in\mathcal{N}\!,t\in\mathcal{M},(\!q,t\!)\ne(\!(t-\!1\!)n +k'\!,t)}\!\left( g_{qt}(l)\right)^{\beta_{qt}}\!\!w_{k'}(l)\!:\beta_{qt}\!\in\![0\!:\!s],k'\!\in\![n]\right\}\!.~~~~~~~~~~\label{align22}
\end{align}
\end{small}
\!\!By (\ref{align22}), we have: \\
1) The product term in the $l$-th row of $\mathbf{U}_{k}$ contains $w_{k}(l)$ with exponent 1, but do not contain $w_{k^{'}}(l),\forall k^{'}\!\ne\!k$. Thus, all the monomial elements in the $l$-th row of $[\mathbf{U}_{1}~\mathbf{U}_{2}~\cdots~ \mathbf{U}_{n}]$ are unique.\\
2) The equations corresponding to $\mathbf{G}_{(i-\!1\!)n+k,i}$ are not contained in the interference alignment relations of (\ref{aligneq2}) for $\mathbf{U}_{k}$, so the monomial elements in the $l$-th row of $\mathbf{U}_{k}$ do not contain $h_{(i-\!1\!)n+k,i}$, $\forall i\!\in\!\mathcal{M}$. It means that all the monomial terms in $\bar{\mathbf{G}}_{(i-\!1\!)n+k,i}\bar{\mathbf{V}}_{\!(i-1\!)n+k}$ are different from those in $\mathbf{U}_{k}$. They are also different from the monomial terms in $\mathbf{U}_{k'}$, $\forall k'\!\ne\! k$, due to $w_{k}(l)$.

Therefore, we can conclude that these $u_s $ vectors in $\mathbf{\Lambda_{j}}$ are independent, and hence $\mathbf{\Lambda_{j}}$ is a full-rank matrix. Taking $s$ to infinity, the scheme achieves a per-receiver DoF of
${\lim_{s \to +\infty} \frac{s^\Gamma}{ s^\Gamma + n(s+1)^{\Gamma}}}= \frac{1}{1+n} = \frac{M}{N+M}$.
Comparing it with the DoF of $\frac{1}{N}$ achieved by TDMA, the per-receiver DoF of the considered downlink channel for $r=1$ is given by $DoF^{d}_1 =\max\left\{\frac{M}{N+M},\frac{1}{N}\right\}$.\\
$ii)$ $r\ge2$:
We first consider interference neutralization enabled by transmitter cooperation. Encode the task output message $\widetilde{W}_{j}$ into $r$ independent streams $x^{p}_{j}$, $p\!\in\![r]$. For better illustration, each stream $x^{p}_{j}$ is given an index $(p\!-\!1)N\!+\!j$. There are a total of $Nr$ (or $Mn$) different streams corresponding to all $N$ messages.
Based on the index order, these $Nr$ different streams can be divided into $N$ groups, each group with $r$ different streams, where the $k$-th group is given by
\begin{equation}
\!\mathcal{Q}_k \!=\! \left\{x^{p}_{j}\!:(p\!-\!1)N\!+\!j\in\left[(k\!-\!1)r\!+\!1\!:kr\right]\right\}, k\in[N].\!
\end{equation}
Since each message exists at $r$ different transmitters, each stream is also owned by $r$ different transmitters. Each group of streams is downloaded in the time division manner. The downlink channel formed by transmitting each group of $r$ streams can be treated as a MISO broadcast channel with perfect transmitter cooperation, whose sum DoF is $r$ \cite{coBC,mimoKuser} achieved by using interference neutralization. Thus, a DoF of $\frac{r}{N}$ is obtained for each receiver in the $r\ge2$ case.

Then, we apply asymptotic interference alignment scheme in the $r\!=\!1$ case. Since $\frac{N}{M}\!=\!n_1$, the messages at each transmitter $i$ are $\big\{\widetilde{W}_{k}\!:k\!\in\![(i\!-\!1)n_1\!+\!1\!:\!in_1]\big\}$ for $r\!=\!1$ and $\big\{\widetilde{W}_{k+\!1}\!:k\!\in\![(i\!-\!1)rn_1\!:\!(irn_1\!-\!1)]\!\!\!\pmod{\!N}\big\}$ for $r\!\ge\!2$. In $r\!\ge\!2$ case, we can let each transmitter only transmit the $n_1$ messages among the total $rn_1$ messages, and different transmitters transmit non-overlapped messages. By doing so, we construct a downlink channel with the same information flow as $r\!=\!\!1$ case. Utilizing the alignment scheme in $r\!=\!1$ case, we thus obtain a per-receiver DoF of $\frac{M}{N+M}$ for $r\!\!\ge\!2$ case. Comparing above two schemes, a Dof of $\max\!\left\{\!\frac{M}{N+M},\frac{r}{N}\!\right\}$ is obtained.

Summarizing the per-receiver DoF for $r\!=\!1$ and $r\!\ge\!2$ cases, we thus prove Lemma \ref{lemma5}.
\subsubsection{Converse proof of Lemma \ref{lemma5}}\label{conversedown}
By using the maximum sum DoF $\min\{M,N\}$ of MISO broadcast channels\cite{misobrocast}, the optimal per-receiver DoF $DoF^{d^{*}}_r$ is upper bounded by $\frac{\min\{M,N\}}{N}$, and the gap between this upper bound and the lower bound in (\ref{dofdp}) satisfies $\frac{DoF^{d^{*}}_r}{DoF^{d}_r} \le 2$.
\vspace{-3mm}
\bibliographystyle{IEEEtran}
\bibliography{refer}
\end{spacing}

\end{document}